\documentclass[11pt]{article}
\usepackage{times,amsmath,amsthm,amssymb,graphicx,hyperref,float,xcolor}
\usepackage{fancyhdr}     
\usepackage{setspace}    
\usepackage{booktabs}    
\usepackage{longtable}   
\usepackage{multirow}
\usepackage{textcomp}    
\usepackage{bm}         
\usepackage{pdfpages}    
\newtheorem{prop}{Properties}
\newtheorem{nt}{Note}

\newtheorem{xmpl}{Example}
\newtheorem{app}{Appendix}
\newtheorem{case}{Case}

\newcommand{\R}{\mathbb{R}}

\newcommand{\ds}{\displaystyle}

\begin{document}
\global\def\refname{{\normalsize \it References:}}
\baselineskip 12.5pt
%
%
%
\title{\LARGE \bf Analytical and Numerical Study of a Convection-Diffusion-Reaction-Source Problem in Multilayered Materials}

\date{}

\author{\hspace*{-10pt}
\begin{minipage}[t]{2.3in} \normalsize \baselineskip 12.5pt
\centerline{GUILLERMO FEDERICO UMBRICHT}
\centerline{Departamento de Matem\'atica, Facultad de Ciencias Empresariales, Universidad Austral}
\centerline{Paraguay 1950, Rosario, Santa Fe, ARGENTINA}
\centerline{Consejo Nacional de Investigaciones Cient{\'i}ficas y T\'ecnicas (CONICET)}
\centerline{Godoy Cruz 2290, CABA, ARGENTINA}
\vspace{0.5cm}
\centerline{DOMINGO ALBERTO TARZIA}
\centerline{Departamento de Matem\'atica, Facultad de Ciencias Empresariales, Universidad Austral}
\centerline{Paraguay 1950, Rosario, Santa Fe, ARGENTINA}
\centerline{Consejo Nacional de Investigaciones Cient{\'i}ficas y T\'ecnicas (CONICET)}
\centerline{Godoy Cruz 2290, CABA, ARGENTINA}
\vspace{0.5cm}
\centerline{DIANA RUBIO}
\centerline{ITECA (UNSAM-CONICET), CEDEMA, ECyT, Universidad
Nacional de General San Mart\'in}
\centerline{25 de mayo y Francia, San Mart\'in, Buenos Aires, ARGENTINA}
\end{minipage}
%
%
\\ \\ \hspace*{-10pt}
\begin{minipage}[b]{6.9in} \normalsize
\baselineskip 12.5pt {\it Abstract:}
In this work, a thermal energy transfer problem in a one-dimensional multilayer body is theoretically analyzed, considering diffusion, advection, internal heat generation or loss linearly dependent on temperature in each layer, as well as heat generation due to external sources. Additionally, the thermal contact resistance at the interfaces between each pair of materials is taken into account. The problem is mathematically modeled, and explicit analytical solutions are derived using Fourier techniques. A convergent finite difference scheme is also formulated to simulate specific cases. The solution is consistent with previous results. A numerical example is provided, demonstrating the coherence between the obtained results and the physical behavior of the problem. This work was recently published for a two-layer body; the generalization to $m$-layer bodies allows for conclusions that enhance the theoretical understanding of heat transfer in multilayer materials and may contribute to improving the thermal design of multilayer engineering systems.
\\ [4mm] {\it Key--Words:}
Heat transfer, Multilayer, Composite materials, Interfacial thermal resistance.\\
%
\end{minipage}
\vspace{-10pt}}

\maketitle

\thispagestyle{empty} \pagestyle{empty}
%
%

\section*{Nomenclature}

\begin{longtable}{p{18mm} c p{120mm} }
\multicolumn{3}{l}{\textbf{Subscripts and Superscripts}}\\
\\
$0$ & --- & initial value\\
$m \, (1,...,M) $ & --- & layer number\\
$M $ & --- & number of body layers\\
$n$ & --- & eigenvalue number\\
$H$ & --- & homogeneous system\\
$\infty$ & --- & stationary state\\
\\
\multicolumn{3}{l}{\textbf{Capital Letters}}\\
\\

$A$ & --- & auxiliary dimensionless parameter\\
$B$ & --- & auxiliary dimensionless parameter\\
$\bar{A}$ & --- & auxiliary temporal function\\
$Bi$ & --- & Biot number \\
$Bi^*$ & --- & auxiliary dimensionless parameter \\
$\bar{Bi}$ & --- & auxiliary dimensionless parameter\\
$C$ & --- & the specific heat at constant pressure $\textbf{[J(kg{$^{\circ}$}C)$^{-1}$]}$\\
$D$ & --- & differential operator $\textbf{[{$^{\circ}$}C\,s$^{-1}$]}$\\
$\bar{D}$ & --- & dimensionless differential operator\\
$K$ & --- & auxiliary dimensionless parameter\\
$L$ & --- & body length $\textbf{[m]}$\\
$N$ & --- & auxiliary dimensionless parameter\\
$Pe$ & --- & Péclet number \\
$P$     & --- & auxiliary function (numerical method) $\textbf{[{$^{\circ}$}C]}$\\
$\mathcal{P}$     & --- & partition (numerical method)\\
$R$ & --- & thermal resistance $\textbf{[m]}$\\
$\bar{R}$ & --- & dimensionless thermal resistance\\
$S$    & --- & auxiliary dimensionless heat source\\
$T$    & --- & temperature field relative to ambient $\textbf{[{$^{\circ}$}C]}$\\
$T_r$ & --- & reference temperature $\textbf{[{$^{\circ}$}C]}$\\
$Z$     & --- & auxiliary parameter (numerical method)\\

\\
\multicolumn{3}{l}{\textbf{Lowercase Letters} }\\
\\
$a$ & --- & auxiliary parameter $\textbf{[{$^{\circ}$}C m$^2$ W$^{-1}$]}$\\
$f$ & --- & dimensionless auxiliary spatial function\\
$g$ & --- & dimensionless auxiliary temporal function\\
$h$     & --- & convection heat transfer coefficient $\textbf{[Wm{$^{-2}$}({$^{\circ}$}C)$^{-1}$]}$\\
$l$ & --- & interface location $\textbf{[m]}$\\
$\bar{l}$ & --- & dimensionless interface location\\
$q$ & --- & auxiliary function\\
$r$ & --- & auxiliary function\\
$s$  & --- & heat source $\textbf{[{$^{\circ}$}C\,s$^{-1}$]}$\\
$\bar{s}$ & --- & dimensionless auxiliary heat source\\
$\widehat{s}$ & --- & dimensionless heat source\\
$t$     & --- & temporary variable $\textbf{[s]}$\\
$t_j$     & --- & particular time (numerical method) $\textbf{[s]}$\\
$x$     & --- & spatial variable $\textbf{[m]}$\\
$x_i$     & --- & particular position (numerical method) $\textbf{[m]}$\\
$y$     & --- & dimensionless spatial variable\\

\\
\multicolumn{3}{l}{\textbf{Greek Letters}}\\
\\

$\alpha$ & --- & thermal diffusivity coefficient $\textbf{[m{$^{2}$}s$^{-1}$]}$\\
$\bar{\alpha}$ & --- & dimensionless thermal diffusivity coefficient\\
$\beta$ & --- & fluid velocity $\textbf{[m\,s$^{-1}$]}$\\
$\nu$ & --- & generation/consumption coefficient $\textbf{[s$^{-1}$]}$\\
$\bar{\nu}$ & --- & dimensionless generation/consumption coefficient\\
$\kappa$ & --- & thermal conductivity coefficient $\textbf{[W(m{$^{\circ}$}C)$^{-1}$]}$\\
$\bar{\kappa}$ & --- & dimensionless thermal conductivity coefficient\\
$\rho$ & --- & density $\textbf{[kg m$^{-3}$]}$\\
$\tau$    & --- & dimensionless temporary variable\\
$\theta$ & --- & dimensionless temperature\\
$\Theta$ & --- & dimensionless auxiliary temperature function\\
$\chi$ & --- & auxiliary dimensionless parameter\\
$\Delta t$     & --- & time discretization step (numerical method) $\textbf{[s]}$\\
$\Delta x$     & --- & spatial discretization step (numerical method) $\textbf{[m]}$\\
$\gamma$ & --- & auxiliary dimensionless parameter\\
$\Gamma$ & --- & auxiliary dimensionless function\\
$\sigma$ & --- & auxiliary dimensionless parameter\\
$\epsilon$     & --- & auxiliary parameter (numerical method)\\
$\varphi$    & --- & auxiliary dimensionless parameter\\
$\psi$     & --- & auxiliary dimensionless parameter\\
$\Psi$     & --- & auxiliary dimensionless parameter\\
$\mu$     & --- & auxiliary dimensionless parameter\\
$\phi$ & --- & auxiliary dimensionless parameter\\
$\xi$ & --- & auxiliary dimensionless parameter\\
$\eta$ & --- & auxiliary dimensionless parameter\\
$\delta$ & --- & auxiliary dimensionless parameter\\
$\iota$     & --- & auxiliary parameter (numerical method)\\
$\lambda$     & --- & dimensionless temporal eigenvalue\\
$\omega$     & --- & dimensionless spatial eigenvalue\\
$\Omega$     & --- & auxiliary parameter (numerical method)\\
$\Lambda$ & --- & auxiliary parameter (numerical method) $\textbf{[W(m{$^{\circ}$}C)$^{-1}$]}$\\
$\upsilon$     & --- & auxiliary parameter (numerical method)\\
$\Pi$     & --- & auxiliary parameter (numerical method) $\textbf{[m$^{-1}$]}$\\
$\zeta$     & --- & auxiliary parameter (numerical method)\\
\end{longtable}

\section{Introduction}
\label{Introduccion} \vspace{-4pt}

The physical and mathematical analysis of mass and heat transfer problems in multilayer composite materials is a topic of extensive current study \cite{Hickson09, Zhou21, Yuan22, Carson22, Yavaraj23}. This interest is primarily due to the direct applications of these problems across various fields of science, engineering, and industry. The breadth of these applications is evident from the abundance of published literature. For example, studies include the growth of brain tumors \cite{Mantzavinos16}, analysis of microelectronic problems \cite{Choobineh15}, thermal conduction in composite materials \cite{Monte00}, drug release analysis in stents \cite{McGinty11}, permeability studies of the skin \cite{Mitragotri11}, moisture analysis in composite tissues \cite{Pasupuleti11}, pollution determination in porous media \cite{Liu98, Liu08}, greenhouse gas emission analysis \cite{Liu09}, lithium-ion cell analysis \cite{Bandhauer11}, innovations in wool cleaning techniques \cite{Caunce08}, and heat conduction through skin analysis \cite{Becker12}, among others.

Mass and/or heat transfer problems in multilayer materials have been analytically addressed using various methods, including recursive image methods \cite{Dias15}, separation of variables \cite{Hickson09, Zhou21, Monte00, Monte02, Ma04, Rubio21}, and solutions involving integral functions such as Laplace and Fourier transforms \cite{Goldner92, Dudko04, Simoes05, Rodrigo16}. Numerical techniques such as the method of fundamental solutions \cite{Johansson09}, finite differences, and finite element methods \cite{Hickson09, Yuan22, Rubio21} have also been employed. A comprehensive and updated review of mass and heat transfer in multilayer materials and the mathematical techniques used can be found in \cite{Monte00, Monte02, Jain21}.

As evident from the previous paragraph, the literature on transport problems in multilayer materials is extensive, but it lacks generality. Most of the cited articles focus solely on diffusion, neglecting other thermal transfer processes. Moreover, many do not consider the thermal contact resistance at the interfaces between each pair of materials. For example, while \cite{Jain21} offers a comprehensive study of heat transfer processes in multilayer materials, it omits the analysis of external heat sources and thermal contact resistance at each interface. Other papers address heat transfer problems in multilayer materials but only under steady-state conditions \cite{Umbricht22, Rubio22, Umbricht22b, Umbricht21, Umbricht20}.

To study more realistic problems, it is essential to understand the full thermal processes, which involve analyzing the influence of external heat generation sources, dissipative terms, and thermal contact resistance. The key physical processes in mass and heat transfer problems in multilayer materials include diffusion, advection, internal heat generation/consumption, and heat generation from external sources. Internal heat generation or consumption rates are often considered proportional to the local temperature. This phenomenon is used in various processes, including perfusion terms in Pennes' biological heat transfer equation \cite{Pennes48}, the fin equations used for segmented multilayer fin analysis \cite{Becker13}, and the kinetics of first-order chemical reactions \cite{Shah16, Esho18}. Advection terms are common in various transfer processes, for example, in flow batteries \cite{Skyllas11}. External heat source terms are useful for modeling processes where heat is delivered to the system through various thermal mechanisms \cite{Kim20}.

In this work, we propose an analytical and numerical study of transient heat transfer in a multilayer body governed by a Convection-Diffusion-Reaction-Source (CDRS) equation. The model considers diffusion, advection, internal heat generation/loss, external heat generation, and thermal contact resistance at the interfaces. An analytical expression for the solution is derived, consistent with previous findings. The existence of infinite eigenvalues is discussed, an orthogonality relation between the spatial functions involved is obtained, and the specific case of two-layer materials is addressed. Additionally, the proposed numerical approach aims to simulate solutions for specific case studies using finite difference methods.

This work was recently published for a two-layer body \cite{Umbricht25}. However, generalizing to 
$m$-layer bodies is necessary since most industrial and natural processes involve composite materials with multiple layers. This type of modeling is crucial for accurately capturing temperature gradients and heat transfer dynamics in more complex systems. Furthermore, multilayer analysis enables more precise and applicable solutions in fields such as materials engineering, biomedicine, and energy industries.

\section{Mathematical Modeling}
\label{Modelo}

The scenario involves the transient thermal energy transfer in a one-dimensional multilayer body. Each layer is assumed to be homogeneous and isotropic. Additionally, heat gain or loss within each layer is considered at a rate proportional to the local temperature, along with advection driven by one-dimensional fluid flow. Heat generation from external sources is also assumed. Thermal runaway phenomena and heat transfer by radiation are neglected.

\begin{figure}[h!]
\begin{center}
\includegraphics[width=0.50\textwidth]{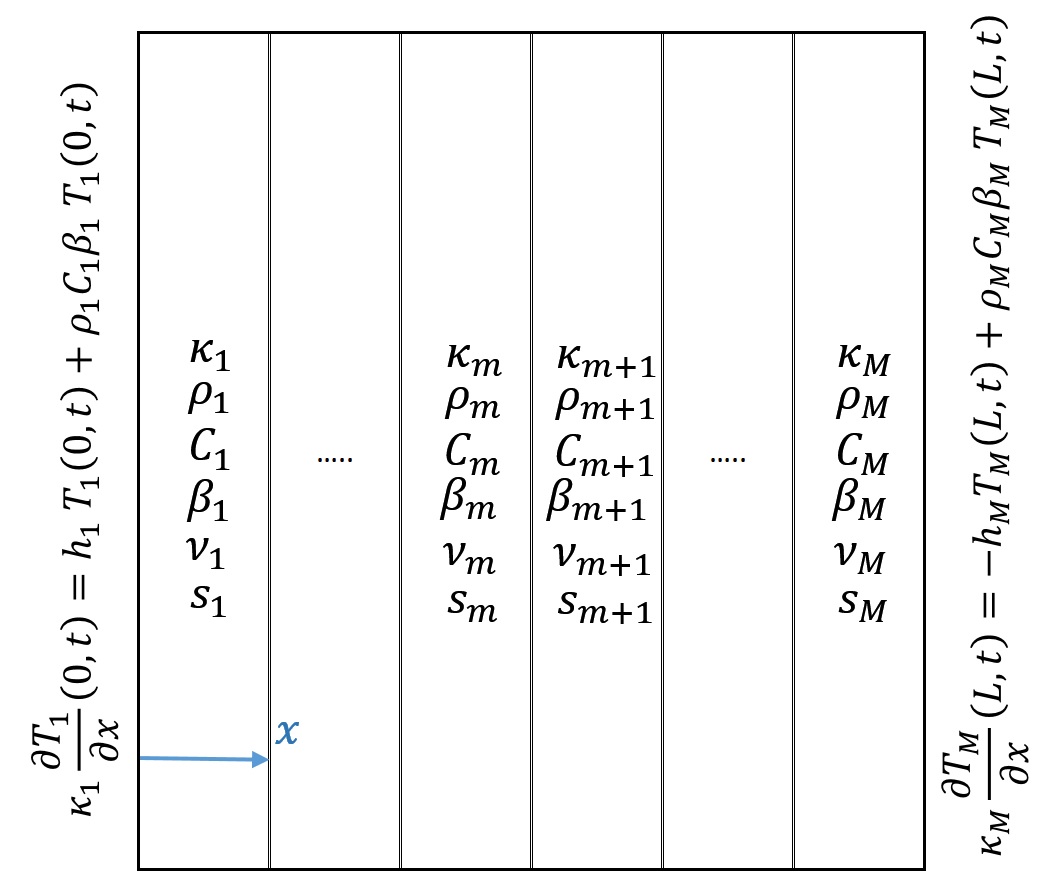}
\caption{General scheme of the problem of interest.}
\label{Esq_Gral}
\end{center}
\end{figure}

The total length of the multilayer body is denoted by $L$. The interface between the $m$-th and $m+1$-th materials is located at position $l_m$ for $m=1,\ldots,M-1$, where $0<l_m<L$. 
In Fig. \ref{Esq_Gral}, a reference diagram is shown, with an arrow indicating the direction of heat flow.

A transient energy conservation equation representing a balance between diffusion, advection, internal heat gain or loss, and heat generation from external sources of a one-dimensional multilayer body can be written as:
\begin{equation}
\label{Ec_Parab}
\rho_m \, C_m \, \dfrac{\partial{T_m}}{\partial{t}}(x,t)= D_m T_m(x,t) + \rho_m \, C_m \, s_m(x,t),  \quad (x,t) \in (l_{m-1},l_{m}) \times \R^+,  \\
\end{equation}
where $D_m$ is a parabolic differential operator that has already been used in other works \cite{Umbricht21b}. This operator is defined as follows for $m=1,...,M$:
\begin{equation}
\label{Operator_D}
D_m T_m(x,t):= \kappa_{m} \, \dfrac{\partial^2{T_m}}{\partial{x^2}}(x,t)-\rho_m \, C_m \,\beta_m \, \dfrac{\partial{T_m}}{\partial{x}} (x,t)+ \rho_m \, C_m \,\nu_m \, T_m(x,t). 
\end{equation}

In the expressions \eqref{Ec_Parab}-\eqref{Operator_D}, commonly referred to as the CDRS equation, the subscripts denote the $m$-th layer of the material, where $x$ and $t$ represent the spatial and temporal variables, respectively. The functions $T_m(x, t)$, satisfying $T_m(x, t) \in C^2(l_{m-1}, l_m) \times C^1(0, +\infty)$, represent the temperature above ambient in the $m$-th layer at position $x$ and time $t$; 
 $\rho_m$ and $C_m$ denote the density and specific heat of each material, respectively.

The first two terms on the right-hand side of equation \eqref{Operator_D} describe heat transfer due to diffusion and advection, while the third term represents internal heat generation or loss proportional to the local temperature. The coefficient $\kappa_m$ denotes the thermal conductivity of the material in each layer, $\beta_m$ represents the flow velocity, and $\nu_m$ corresponds to the coefficient relating the rate of internal heat generation or loss to the local temperature. The differentiable functions $s_m$, given in \eqref{Ec_Parab}, model an external heat source acting on the body. All material properties are assumed to be temperature-independent. A similar equation can be applied to model the concentration field in a one-dimensional mass transfer problem \cite{Kim20}.

Heat is generated due to external sources and within each layer at a rate proportional to the local temperature. Heat transfer within the body occurs via diffusion and advection, driven by a one-dimensional fluid flow imposed in each layer, flowing from left to right. Each layer is characterized by distinct thermal properties, flow velocity, and internal heat generation rate.

General convective boundary conditions are assumed at the left and right boundaries, respectively. These conditions represent a balance between two factors: convective heat transfer between the body and the surroundings, and diffusion and advection into and out of the body. Note that while advection transfers energy from the surroundings to the first layer, it also removes energy from the last layer to the surroundings.
\begin{equation}
\label{Cond_Borde}
\begin{cases}
\kappa_1 \, \dfrac{\partial{T_1}}{\partial{x}}(x,t)=h_1 \,T_1(x,t) + \rho_1  C_1  \beta_1 \,T_1(x,t) , & \, x=0, \,\,\, t \in \R^+, 
\vspace{0.2cm}
\\
\kappa_M \, \dfrac{\partial{T_M}}{\partial{x}}(x,t)=-h_M \,T_M(x,t) + \rho_M  C_M  \beta_M \,T_M(x,t), & \, x=L, \,\,\, t \in \R^+,
\end{cases} 
\end{equation}
where $h_m$ for $m=1,\ldots,M$ denotes the convection heat transfer coefficient.

Additionally, the temperature discontinuity at each interface is taken into account due to the thermal contact resistance at the junction of each pair of materials. Thus, for $m=1,2,...,m-1$, it holds that:
\begin{equation}
\label{saltotermico}
T_{m+1}(x,t)=T_m(x,t)+ a_m \, \kappa_m \, \dfrac{\partial{T_m}}{\partial{x}} (x,t), \qquad  x=l_m, \, \, t \in \R^+. 
\end{equation}
where $a_m$ is a constant that depends on the physical configuration of the surface in thermal contact, and $a_m \, \kappa_m$ represents the thermal contact resistance at the $m$-th interface, which, for simplicity, will hereafter be denoted as $R_m$. Additionally, by applying energy conservation, this implies continuity of the heat flux across each interface. That is to say,
\begin{equation}
\begin{split}
\label{Cond_Interf}
& \kappa_{m+1}  \dfrac{\partial{T_{m+1}}}{\partial{x}}(x,t)- \rho_{m+1} \, C_{m+1} \, \beta_{m+1} \, T_{m+1}(x,t)  \\
& =\kappa_m \dfrac{\partial{T_m}}{\partial{x}}(x,t)- \rho_m \, C_m \, \beta_m \, T_m(x,t), \qquad x=l_m, \, \, t \in \R^+, 
\end{split}
\end{equation}

Finally, an initial spatial distribution of temperature in each layer is assumed. This implies the following conditions
\begin{equation}
\label{Cond_Inicial}
T_m(x,t)=T_{m,0} (x),  \qquad x \in \left[ l_{m-1},l_{m} \right], \,\,\, t=0. 
\end{equation} 

\begin{nt}
The problem described by equations \eqref{Ec_Parab}-\eqref{Cond_Inicial} is analyzed at a macroscopic scale, as the findings may not hold true for other scales. This is largely because the thermophysical properties of interfaces between materials, as well as their effects, can vary considerably depending on the scale. For example, at the nanoscale, the one-dimensional heat transfer problem between two layers cannot be adequately solved using the methods outlined in this work. At that scale, alternative approaches, such as non-equilibrium molecular dynamics simulations or non-equilibrium Green’s function methods based on interatomic potentials, are required. This nanoscale issue is particularly important in the study of interface nanodevices and has recently been explored by several researchers \cite{Yang19,Li23,Yang23,Yang24} for different materials, including graphene-silver, graphene-gold, graphene-silicon, and graphene-copper.
\end{nt} 

In the next section we obtain an explicit analytical solution to the problem we have just described given by the equations \eqref{Ec_Parab}-\eqref{Cond_Inicial}.

\section{Analytical Solution} \label{Solucion_analitica}

The transient heat transfer problem to be solved is defined by the equations \eqref{Ec_Parab}-\eqref{Cond_Inicial}. To simplify the approach, the expressions are non-dimensionalized by introducing the following parameters for $m=1,2,...,M$,
\begin{equation}
\label{Aux0}
\begin{cases} 
y=\dfrac{x}{L}, \,\,\, \bar{l}_m=\dfrac{l_m}{L}, \,\,\, \bar{R}_m=\dfrac{R_m}{L}, \,\,\,\tau=\dfrac{\alpha_M}{L^2} \,t, \,\,\, \theta_m=\dfrac{T_m}{T_{r}}, \,\,\, \bar{\alpha}_m=\dfrac{\alpha_m}{\alpha_M}, \vspace{0.2cm}\\ {Pe}_m=\dfrac{L}{\alpha_M} \, \beta_m,  \,\,\, 
 \bar{\nu}_m=\dfrac{L^2}{\alpha_M} \,\nu_m, \,\,\, \bar{s}_m= \dfrac{L^2}{T_{r} \, \alpha_M} \, s_m, \,\,\, \bar{\kappa}_m=\dfrac{\kappa_m}{\kappa_M}, \vspace{0.2cm} \\ {Bi}_1=\dfrac{L}{\kappa_M} \, h_1, \,\,\, {Bi}_M=\dfrac{L}{\kappa_M} \, h_M,
\end{cases}
\end{equation}
where $\alpha_m=\dfrac{\kappa_m}{\rho_m C_m}$ represents the termal diffusivity coefficient of de $m$-th material, $Pe_m$ and $Bi_m$ denote the dimensionless Péclet and Biot numbers, respectively, and the parameter $T_r$ represents any reference temperature. This change of variables is applied to equations \eqref{Ec_Parab}-\eqref{Cond_Inicial}, resulting in the following dimensionless
system:
\begin{equation}
\label{SisT_adimen}
\begin{cases} 
\dfrac{\partial{\theta_m}}{\partial{\tau}}(y,\tau)= \bar{D}_m \theta_m (y,\tau) + \bar{s}_m(y,\tau), \, & (y,\tau) \in (\bar{l}_{m-1},\bar{l}_{m}) \times \R^+, \vspace{0.1cm}\\
\dfrac{\partial{\theta_1}}{\partial{y}}(y,\tau)=  {{Bi}_1}^* \, \theta_1(y,\tau) , \, & y=0 , \, \tau \in \R^+, \vspace{0.1cm}\\
\dfrac{\partial{\theta_M}}{\partial{y}}(y,\tau)=  {{Bi}_M}^* \, \theta_M(y,\tau) , \, & y=1 , \, \tau \in \R^+, \\
\theta_{m+1}(y,\tau) = \theta_m(y,\tau) + \bar{R}_m \, \dfrac{\partial{\theta_m}}{\partial{y}}(y,\tau) , \, & y=\bar{l}_m , \, \tau \in \R^+, \\
\dfrac{\partial{\theta_{m+1}}}{\partial{y}}(y,\tau)= \gamma_m \, \theta_m(y,\tau) + \sigma_m \, \dfrac{\partial{\theta_m}}{\partial{y}}(y,\tau), \, & y=\bar{l}_m , \, \tau \in \R^+, \\
\theta_m(y,\tau)=\theta_{m,0} (y), \, & y \in \left[\bar{l}_{m-1},\bar{l}_{m}\right], \, \tau=0,
\end{cases}
\end{equation}
where
\begin{equation}
\label{Oper_adimen}
\bar{D}_m \theta_m (y,\tau)=\bar{\alpha}_m \, \dfrac{\partial^2{\theta_m}}{\partial{y^2}}(y,\tau) - Pe_m \,\dfrac{\partial{\theta_m}}{\partial{y}}(y,\tau) + \bar{\nu}_m \, \theta_m(y,\tau),  
\end{equation}
and 
\begin{equation}
\begin{cases} 
\label{Par_nuevos}
{{Bi}_1}^*=\dfrac{Pe_1}{\bar{\alpha}_1}+\dfrac{{Bi}_1}{\bar{\kappa}_1},  \,\,
{{Bi}_M}^*= Pe_M - {Bi}_M,\vspace{0.3cm}\\
\gamma_m= \dfrac{Pe_{m+1}}{\bar{\alpha}_{m+1}}- \dfrac{Pe_{m}}{\bar{\alpha}_{m}}\, \dfrac{\bar{\kappa}_m}{\bar{\kappa}_{m+1}},\,\, 
\sigma_m=\dfrac{\bar{\kappa}_m}{\bar{\kappa}_{m+1}}+ \bar{R}_m \, \dfrac{Pe_{m+1}}{\bar{\alpha}_{m+1}}.
\end{cases}
\end{equation}

Then, the advective term is removed from equation \eqref{Oper_adimen}by applying a substitution that can be interpreted as a change in the coordinate system. This
transformation effectively shifts the system into a reference frame moving
with the fluid velocity. Similar coordinate system changes have been used in
the literature to address various situations. For instance, see \cite{Basha93,Bharati17,Das17,Sanskrityayn17}. The proposed substitution in this case is:
\begin{equation}
\label{Def_w}
\theta_m(y,\tau)=\exp\left(\chi_m\,y\right) \, \Theta_m(y,\tau),  \qquad  (y,\tau) \in [\bar{l}_{m-1},\bar{l}_m] \times \R^+,
\end{equation}
where
\begin{equation}
\label{chis}
\chi_m= \dfrac{Pe_m}{2 \,\bar{\alpha}_m}.
\end{equation}
The change of variables \eqref{Def_w}-\eqref{chis} is applied to equations \eqref{SisT_adimen}-\eqref{Par_nuevos} leading to the following system
\begin{equation}
\label{Sist_w1} 
\begin{cases} 
\dfrac{\partial{\Theta_m}}{\partial{\tau}}(y,\tau)=\bar{\alpha}_m \, \dfrac{\partial^2{\Theta_m}}{\partial{y^2}}(y,\tau) + \psi_m \, \Theta_m(y,\tau) + \widehat{s}_m(y,\tau), \, & (y,\tau) \in (\bar{l}_{m-1},\bar{l}_{m}) \times \R^+, \\
\dfrac{\partial{\Theta_1}}{\partial{y}}(y,\tau)=  \bar{{Bi}}_1 \, \Theta_1(y,\tau) , \, & y=0 , \, \tau \in \R^+, \vspace{0.1cm}\\
\dfrac{\partial{\Theta_M}}{\partial{y}}(y,\tau)=  \bar{{Bi}}_M \, \Theta_M(y,\tau) , \, & y=1 , \, \tau \in \R^+, \\
\Theta_{m+1}(y,\tau) = \phi_m \, \Theta_m(y,\tau) + \mu_m \, \dfrac{\partial{\Theta_m}}{\partial{y}}(y,\tau) , \, & y=\bar{l}_m , \, \tau \in \R^+, \\
\dfrac{\partial{\Theta_{m+1}}}{\partial{y}}(y,\tau)= \eta_m \, \Theta_m(y,\tau) + \varphi_m \, \dfrac{\partial{\Theta_m}}{\partial{y}}(y,\tau), \, & y=\bar{l}_m , \, \tau \in \R^+, \\
\Theta_m(y,\tau)=\Theta_{m,0} (y), \, & y \in \left[\bar{l}_{m-1},\bar{l}_{m}\right], \, \tau=0,
\end{cases}
\end{equation}
where
\begin{equation}
\label{Aux1}
\begin{cases} 
\psi_m= \bar{\nu}_m - \bar{\alpha}_m \, \chi_m^2, \,\,\, \widehat{s}_m(y,\tau)= \bar{s}_m(y,\tau) \, \exp\left(-\chi_m \, y \right), 
 \,\,\,
\bar{{Bi}}_1 = {{Bi}_1}^* -\chi_1, \\  \bar{{Bi}}_M = {{Bi}_M}^*-\chi_M, \,\,\,  \phi_m= \xi_m \,\delta_m , \,\,\, \mu_m= \xi_m\,\bar{R}_m ,\,\,\,\varphi_m=\xi_m \, \left(\sigma_m - \bar{R}_m \,\chi_{m+1} \right),\\ \eta_m=\xi_m \, \left(\gamma_m+ \sigma_m \,\chi_m - \delta_m\,\chi_{m+1}   \right), \,\,\, 
  \xi_m= \exp\left(\, \bar{l}_m \, (\chi_m-\chi_{m+1})\right),  \\ \delta_m=1+ \bar{R}_m \, \chi_m,  \,\,\, \Theta_{m,0} (y)=\theta_{m,0} (y) \, \exp\left(-\chi_m \, y \right).
\end{cases}
\end{equation}

The solution of the non-homogeneous system \eqref{Sist_w1}-\eqref{Aux1} is obtained using classical techniques for solving partial differential equations. 

First, the associated homogeneous system is solved by applying the method of separation of variables, yielding a solution of the following form: 
\begin{equation}
\label{Soluciongeneralhomogeneo}
\Theta_m^H(y,\tau)= \ds \sum_{n=1}^{\infty} {f_{m,n} (y) \, \left(K_n \,\exp(-\lambda_n^2 \, \tau)\right)}, \qquad  (y,\tau) \in (\bar{l}_{m-1},\bar{l}_m) \times \R^+,
\end{equation}
where $f_{m,n}$ is a sequence of functions depending solely on the dimensionless spatial variable $y$ for each $m = 1, \ldots, M$, the sequence $K_n$ is associated with the initial temperature distribution, and $\lambda_n$ are the temporal eigenvalues. Details of the solution to the associated homogeneous system can be found in the Appendix \ref{SHS}. The discussion regarding the existence of infinitely many real solutions $\lambda_n$ to the eigenvalue equation is presented in the Appendix \ref{Autovalores}. Furthermore, the orthogonality relation of the functions $f_{m,n}$, which will be used to solve the non-homogeneous problem, is derived in the Appendix \ref{Ortogonalidad}.

The solution of \eqref{Sist_w1}-\eqref{Aux1} is obtained using the Fourier technique, yielding:  
\begin{equation}
\Theta_m(y,\tau)= \ds \sum_{n=1}^{\infty} {f_{m,n} (y) \, \bar{A}_{m,n}(\tau)}, \quad  (y,\tau) \in [\bar{l}_{m-1},\bar{l}_m] \times \R^+,
\end{equation}
where $\bar{A}_{m,n}(\tau)$ is a sequence of functions depending on the dimensionless temporal variable $\tau$ for each $m = 1, \ldots, M$. Details of its derivation can be found in the Appendix \ref{SNHS}.

Finally, substituting the expression for $\Theta_m$ into \eqref{Def_w}, the solution to the dimensionless problem of interest \eqref{SisT_adimen}-\eqref{Par_nuevos} is obtained as:  
\begin{equation}
\theta_m(y,\tau)= \ds \sum_{n=1}^{\infty} {\exp\left(\chi_m\,y\right) \, f_{m,n} (y) \, \bar{A}_{m,n}(\tau)}, \quad  (y,\tau) \in [\bar{l}_{m-1},\bar{l}_m] \times \R^+.
\end{equation}

\section{Particular case (two-layer material)} \label{Caso_Bicapa}

In this section, the particular case for a bilayer material is derived from the result obtained in this work. The solution is given by:

\begin{equation}
\begin{cases}
\label{SolNoHomegeneo2bis}
\theta_1(y,\tau)= \ds \sum_{n=1}^{\infty} {\exp\left(\chi_1\,y\right) \,\bar{A}_{1,n}(\tau) \, f_{1,n} (y) }, \quad  (y,\tau) \in [0,\bar{l}] \times \R^+,  \\
\theta_2(y,\tau)= \ds \sum_{n=1}^{\infty} {\exp\left(\chi_2\,y\right) \,\bar{A}_{2,n}(\tau) \, f_{2,n} (y) }, \quad  (y,\tau) \in [\bar{l},1] \times \R^+.
\end{cases}
\end{equation}
where  
\begin{equation}
\label{chisbis}
\chi_1= \dfrac{Pe_1}{2 \,\bar{\alpha}}, \qquad  \qquad \chi_2= \dfrac{Pe_2}{2 },
\end{equation}
for
\begin{equation}
\label{Aux0bis}
\bar{\alpha}=\dfrac{\alpha_1}{\alpha_2}, \qquad \qquad {Pe}_1=\dfrac{L}{\alpha_2} \, \beta_1, \qquad \qquad {Pe}_2=\dfrac{L}{\alpha_2} \, \beta_2.
\end{equation}

The functions $f_{1,n}$ and $f_{2,n}$ of \eqref{SolNoHomegeneo2bis} are given by the following expressions:
\begin{equation}
\label{fs2bis}
\begin{cases}
f_{1,n}(y)= \cos (\omega_{1,n} \, y) + \dfrac{\bar{{Bi}}_1}{\omega_{1,n}} \sin (\omega_{m,n}\, y), \qquad  y \in [0,\bar{l}]. \vspace{0.2cm} \\
f_{2,n}(y)= A_{n}  \cos (\omega_{2,n} \, y) + B_{n}  \sin (\omega_{2,n}\, y), \qquad  y \in [\bar{l},1],
\end{cases}
\end{equation}
where
\begin{equation}
\label{Omegasbis2}
\begin{cases}
\omega_{1,n}=\omega_{1,n}(\lambda_n)=\sqrt{\dfrac{\lambda_n^2+\psi_1}{\bar{\alpha}}}=\sqrt{\dfrac{\lambda_n^2+\bar{\nu}_1-\bar{\alpha} \chi_1^2}{\bar{\alpha}}}=\sqrt{\dfrac{\lambda_n^2+\bar{\nu}_1- \frac{Pe_1^2}{4\bar{\alpha}}}{\bar{\alpha}}},\\
\omega_{1,n}=\omega_{m,n}(\lambda_n)=\sqrt{\lambda_n^2+\psi_2}=\sqrt{\lambda_n^2+\bar{\nu}_2-\chi_2^2}=\sqrt{\lambda_n^2+\bar{\nu}_2- \frac{Pe_2^2}{4}}
\end{cases}
\end{equation}
and
\begin{equation}
\label{Anbis}
\begin{split}
A_{n}&= \dfrac{\sin(\omega_{1,n} \, \bar{l})}{\cos(\omega_{2,n} \, \bar{l})} \left(\phi \,\dfrac{\bar{{Bi}}_1}{\omega_{1,n}}-\mu \, \omega_{1,n} \right)
  + \dfrac{\cos(\omega_{1,n} \, \bar{l})}{\cos(\omega_{2,n} \, \bar{l})} \left(\phi  +\mu \, \bar{{Bi}}_1\right) \\  & - \tan (\omega_{2,n} \, \bar{l}) \, B_{n}   
\end{split}
\end{equation}
\begin{equation}
\label{Bnbis}
\begin{split}
 B_{n} = & \sin(\omega_{2,n} \, \bar{l}) \left[\sin(\omega_{1,n} \, \bar{l}) \left(\phi \, \dfrac{\bar{{Bi}}_1}{\omega_{1,n}} - \mu \, \omega_{1,n} \right) +\cos(\omega_{1,n} \, \bar{l}) \left(\phi + \mu \, \bar{{Bi}}_1\right)\right] 
 \\
 +& \dfrac{\cos(\omega_{2,n} \, \bar{l})}{\omega_{2,n}}  \left[\sin(\omega_{1,n} \, \bar{l})\left(\eta \, \dfrac{\bar{{Bi}}_1}{\omega_{1,n}} - \varphi \, \omega_{1,n} \right) + \cos(\omega_{1,n} \, \bar{l}) \left(\eta +\varphi \, \bar{{Bi}}_1 \right)\right] 
\end{split}
\end{equation}
The eigenvalues $\lambda_n$ are the infinite solutions of the transcendental eigenvalue equation given by:
\begin{equation}
\label{Tangbis2}
\tan (\omega_{2,n}(\lambda_n))=\dfrac{\omega_{2,n} (\lambda_n) \, B_{n}- \bar{{Bi}}_2 \, A_{n}}{\bar{{Bi}}_2 \, B_{n}+\omega_{2,n}(\lambda_n)\,  A_{n}},
\end{equation}
where 

\begin{equation}
\begin{cases}
\label{Par_nuevosbis}
\bar{l}=\dfrac{l}{L}, \,\,\, \bar{R}=\dfrac{R}{L}, \,\,\, \bar{{Bi}}_1 = \dfrac{Pe_1}{\bar{\alpha}}+\dfrac{h_1 \, L}{\kappa_1}-\chi_1, \,\,\, 
{{Bi}_2}^*= Pe_2 - \dfrac{h_2 \, L}{\kappa_2}-\chi_2,\,\\
\gamma= Pe_{2}-\dfrac{\bar{\kappa}}{\bar{\alpha}}-Pe_{1}, \,\,\, 
\sigma=\bar{\kappa}+ \bar{R}\,Pe_2, \,\,\,
\phi= \xi \,\delta , \,\,\, \mu= \xi\,\bar{R} ,\,\,\,\varphi=\xi \, \left(\sigma - \bar{R} \,\chi_{2} \right),\\ \delta=1+ \bar{R} \, \chi_1, \,\,\, \eta=\xi \, \left(\gamma+ \sigma \,\chi_1 - \delta\,\chi_{2}\right), \,\,\, 
  \xi= \exp\left(\, \bar{l} \, (\chi_1-\chi_{2})\right),  \\   
\psi_1= \dfrac{L^2}{\alpha_2} \,\nu_1 - \bar{\alpha} \, \chi_1^2, \,\,\, \psi_1= \dfrac{L^2}{\alpha_2} \,\nu_2 - \chi_2^2. 
\end{cases}
\end{equation}

Finally, the functions $\bar{A}_{1,n}$ and $\bar{A}_{2,n}$ of \eqref{SolNoHomegeneo2bis} are given by the following expressions:
\begin{equation}
\label{solEDObis}
\begin{cases}
\bar{A}_{1,n}(\tau)=\exp\left((\psi_1-\bar{\alpha} \, \omega^2_{1,n})\, \tau\right) \left[K_n + \ds \int_0^\tau S_{1,n} (s) \, \exp\left((\bar{\alpha} \, \omega^2_{1,n}-\psi_1)\, s\right)  \,ds\right], \\ 
\bar{A}_{2,n}(\tau)=\exp\left((\psi_2- \omega^2_{2,n})\, \tau\right) \left[K_n + \ds \int_0^\tau S_{2,n} (s) \, \exp\left((\omega^2_{2,n}-\psi_2)\, s\right)  \,ds\right],
\end{cases}
\end{equation}

where
\begin{equation}
\label{Gsbis}
S_{1,n}(\tau)=\dfrac{\ds \int_{0}^{\bar{l}} \widehat{s}_1(y,\tau) \,  f_{1,n} (y) \, dy}{\ds \int_{0}^{\bar{l}} \left[f_{1,n} (y)\right]^2 \, dy}, \qquad S_{2,n}(\tau)=\dfrac{\ds \int_{\bar{l}}^{1} \widehat{s}_2(y,\tau) \,  f_{2,n} (y) \, dy}{\ds \int_{\bar{l}}^{1} \left[f_{2,n} (y)\right]^2 \, dy},
\end{equation}

for 
\begin{equation}
\label{Fuenteenseriebis}
\begin{cases}
\widehat{s}_1(y,\tau)= \ds \sum_{n=1}^{\infty}  { S_{1,n}(\tau) \, f_{1,n} (y) }, \quad  (y,\tau) \in [0,\bar{l}] \times \R^+,\\
\widehat{s}_2(y,\tau)= \ds \sum_{n=1}^{\infty}  { S_{2,n}(\tau) \, f_{2,n} (y) }, \quad  (y,\tau) \in [\bar{l},1] \times \R^+
\end{cases}
\end{equation}
and

\begin{equation}
\label{CIbis}
K_n= \dfrac{\dfrac{\varphi\, \phi-\eta \, \mu}{\bar{\alpha}} \ds \int_{0}^{\bar{l}} \theta_{1,0}(y)\, f_{1,n}(y)\,\exp\left(-\chi_1\,y\right)\, dy 
+\ds \int_{\bar{l}}^1 \theta_{2,0}(y)\, f_{2,n}(y)\, \exp\left(-\chi_2\,y\right) dy}{\dfrac{\varphi\, \phi-\eta \, \mu}{\bar{\alpha}} \ds \int_{0}^{\bar{l}} [f_{1,n}(y)]^2 \, \exp\left(-\chi_1\,y\right)\, dy 
+\ds \int_{\bar{l}}^1 [f_{2,n}(y)]^2 \, \exp\left(-\chi_2\,y\right)\, dy}.
\end{equation}

The solution derived in this paper proves to be robust, since for the particular case of a bilayer body, it coincides with the solution provided by the authors in \cite{Umbricht25}.

\section{Consistency validation} \label{Consistencia}

There are several ways to analyze the consistency of the solution obtained with those existing in the literature. In \cite{Jain21}, the authors consider a situation similar to the one addressed here, but with simpler characteristics that are of special interest for carrying out this analysis. In this article, external heat sources are neglected and thermal contact resistance at the interface is not considered. We are interested in seeing that, under these assumptions, both solutions are equal.

For this particular case, since there are no external heat sources, we have $s_m=0$. Furthermore, since thermal resistance at each interface is neglected, $R_m=0$.

Because the external sources are null $(s_m=0)$, the problem is reduced to considering the solution of the associated homogeneous system given by
\begin{equation}
\label{Sol2}
\Theta_m^H(y,\tau)= \ds \sum_{n=1}^{\infty} { K_n \left[A_{m,n}  \cos (\omega_{m,n} \, y) + B_{m,n}  \sin (\omega_{m,n}\, y) \right] \exp(-\lambda_n^2 \, \tau)}, 
\end{equation}
where 
\begin{equation}
\label{Omegas2}
\omega_{m,n}=\omega_{m,n}(\lambda_n)=\sqrt{\dfrac{\lambda_n^2+\psi_m}{\bar{\alpha}_m}}=\sqrt{\dfrac{\lambda_n^2+\bar{\nu}_m-\bar{\alpha}_m \chi_m^2}{\bar{\alpha}_m}}=\sqrt{\dfrac{\lambda_n^2+\bar{\nu}_m- \frac{Pe_m^2}{4\bar{\alpha}_m}}{\bar{\alpha}_m}}
,
\end{equation}
wich
$A_{1,n}=1$, $B_{1,n}=\frac{\bar{{Bi}}_1}{\omega_{1,n}}$ and for $m=2,...,M-1$
\begin{equation}
\label{An2bis}
\begin{split}
A_{m+1,n}&= \dfrac{\sin(\omega_{m,n} \, \bar{l}_m)}{\cos(\omega_{m+1,n} \, \bar{l}_m)} \left(\phi_m \, B_{m,n}-\mu_m \, \omega_{m,n} \, A_{m,n}\right)
\\  &   + \dfrac{\cos(\omega_{m,n} \, \bar{l}_m)}{\cos(\omega_{m+1,n} \, \bar{l}_m)} \left(\phi_m \, A_{m,n} +\mu_m \, \omega_{m,n} \, B_{m,n}\right) - \tan (\omega_{m+1,n} \, \bar{l}_m) \, B_{m+1,n} 
\end{split}
\end{equation}
and
\begin{equation}
\label{Bn2bis}
\begin{split}
 B_{m+1,n} = & \sin(\omega_{m+1,n} \, \bar{l}_m) \left[\sin(\omega_{m,n} \, \bar{l}_m) \left(\phi_m \, B_{m,n} - \mu_m \, \omega_{m,n} \, A_{m,n}\right)  \right] \\
 +& \sin(\omega_{m+1,n} \, \bar{l}_m) \left[ \cos(\omega_{m,n} \, \bar{l}_m) \left(\phi_m \, A_{m,n}+ \mu_m \, \omega_{m,n} \, B_{m,n}\right)\right] \\
 +& \dfrac{\cos(\omega_{m+1,n} \, \bar{l}_m)}{\omega_{m+1,n}}  \left[\sin(\omega_{m,n} \, \bar{l}_m)\left(\eta_m \, B_{m,n} - \varphi_m \, \omega_{m,n} \, A_{m,n}\right) \right] \\
+& \dfrac{\cos(\omega_{m+1,n} \, \bar{l}_m)}{\omega_{m+1,n}}  \left[ \cos(\omega_{m,n} \, \bar{l}_m) \left(\eta_m \, A_{m,n}+\varphi_m \, \omega_{m,n} \, B_{m,n} \right)\right].
\end{split}
\end{equation}
The eigenvalues $\lambda_n$ are the infinitily many solutions of the equation
\begin{equation}
\label{Tang2}
\tan (\omega_{M,n})=\dfrac{\omega_{M,n} \, B_{M,n}- \bar{{Bi}}_M \, A_{M,n}}{\bar{{Bi}}_M \, B_{M,n}+\omega_{M,n}\,  A_{M,n}},
\end{equation}
Finally, $K_n$ is determined from the initial conditions using the orthogonality principle discussed in the Appendix \ref{Ortogonalidad}.
\begin{equation}
\label{CI2}
K_n= \dfrac{\ds \sum_{m=1}^{M} \dfrac{\Psi_{m}}{\bar{\alpha}_m} \, \ds \int_{\bar{l}_{m-1}}^{\bar{l}_m} \Theta_{m,0}(y)\, f_{m,n}(y) \, dy }{\ds \sum_{m=1}^{M}  \dfrac{\Psi_{m}}{\bar{\alpha}_m} \ds \int_{\bar{l}_{m-1}}^{\bar{l}_m}  [f_{m,n}(y)]^2 \, dy }.
\end{equation}
The only remaining step is to impose the absence of contact resistance at the interface. To do this, it is necessary to evaluate the equations \eqref{Sol2}-\eqref{CI2} at $R=0$. Then the parameters that are modified from these changes are:
\begin{equation}
\label{Aux1bis2}
\phi_m= \xi_m , \quad \mu_m= 0 ,\quad  \varphi_m=\xi_m \,\dfrac{\bar{\kappa}_m}{\bar{\kappa}_{m+1}}  \quad  \eta_m=\xi_m \, \left(\gamma_m+ \dfrac{\bar{\kappa}_m}{\bar{\kappa}_{m+1}}\,\chi_m - \chi_{m+1}   \right).
\end{equation}
In summary, when we examine the solution (derived in this article) for the
specific case of transient heat transfer with no thermal sources and neglecting
contact resistance at the interface, it is found that the solution satisfies the
conditions provided by the authors in \cite{Jain21}.

What has been presented in this section aims to demonstrate that the more complex model is consistent with the known solutions in the literature for simpler, specific cases.

\section{Numerical Modelling}

The analytical solution of this type of problems has a high numerical burden, which makes it complex to obtain temperature profiles for specific cases. Because of this, the problem in question is usually modeled using some numerical method that allows graphing different temperature profiles and obtaining information from them.

The finite difference method is often an effective tool for evolutionary heat transfer problems. When dealing with multilayer bodies, the junction of each pair of materials often poses a challenge, especially if there is no temperature continuity. Some authors have addressed this situation by incorporating virtual or artificial layers; see, for example, \cite{Yuan22}.

In this work, we propose an explicit second-order finite difference method that employs both a forward-in-time scheme and a centered-in-space scheme, along with specific adaptations at the boundaries and interfaces. At the right boundary, we apply backward differences, while at the left boundary, we utilize forward differences. For the interface, the approach involves either forward or backward differences depending on whether the left or right material is being considered. 

To implement the numerical method, $M$ uniform two-dimensional partitions are defined in the spatial variable $x$ and the time variable $t$ as a discrete set $\mathcal{P}$ that satisfies:
\begin{equation}
\label{particion1}
\begin{cases}
\mathcal{P}_1=\{ (x_i,t_j)/ \, i=1,2,...,n_{l_1} ;\, j=1,2,...,J;\, x_i \in \mathcal{P}_x^1,\, t_j \in \mathcal{P}_t \}, \\
\mathcal{P}_2=\{ (x_i,t_j)/ \, i=n_{l_1},n_{l_1+1},...,n_{l_2} ;\, j=1,2,...,J;\, x_i \in \mathcal{P}_x^2,\, t_j \in \mathcal{P}_t \}, \\
\mathcal{P}_3=\{ (x_i,t_j)/ \, i=n_{l_2},n_{l_2+1},...,n_{l_3} ;\, j=1,2,...,J;\, x_i \in \mathcal{P}_x^3,\, t_j \in \mathcal{P}_t \}, \\
\quad \vdots \qquad \qquad\qquad \qquad\qquad\qquad\vdots\\
\mathcal{P}_M=\{ (x_i,t_j)/ \, i=n_{l_{M-1}},...,n_{l_M} ;\, j=1,2,...,J;\, x_i \in \mathcal{P}_x^M,\, t_j \in \mathcal{P}_t \},
\end{cases}
\end{equation}
where
\begin{equation}
\label{particion2}
\begin{cases}
\mathcal{P}_x^1=\{ x_1< \cdots < x_i< \cdots <x_{n_{l_1}}, \,\,\, x_i=(i-1) \Delta x, \,\,\, i=1,2,...,n_{l_1}\}, \\
\mathcal{P}_x^2=\{ x_{n_{l_1}}<  \cdots < x_i< \cdots <x_{n_{l_2}}, \,\,\, x_i=(i-1) \Delta x, \,\,\, i=n_{l_1},...,n_{l_2}\}, \\
\mathcal{P}_x^3=\{ x_{n_{l_2}}<  \cdots < x_i< \cdots <x_{n_{l_3}}, \,\,\, x_i=(i-1) \Delta x, \,\,\, i=n_{l_2},...,n_{l_3}\}, \\
\quad \vdots \qquad \qquad\qquad \qquad\qquad\qquad\vdots \\
\mathcal{P}_x^M=\{ x_{n_{l_{M-1}}}<  \cdots < x_i< \cdots <x_{n_{l_{M}}}, \,\,\, x_i=(i-1) \Delta x, \,\,\, i=n_{l_{M-1}},...,n_{l_{M}}\}
\end{cases}
\end{equation}
and
\begin{equation}
\label{particion3}
\mathcal{P}_t=\{ t_1< t_2< \cdots < t_j< \cdots <t_M, \,\,\, t_j=(j-1) \Delta t, \,\,\, j=1, 2,...,J\}.
\end{equation}

Specifically, $\mathcal{P}_x^m$ with $m=1,2,...,M$ is the partition of the spatial variable $x$ associated with $T_m$, and $\mathcal{P}_t$ is the corresponding partition associated with the time variable $t$. The values of $\Delta x$ and $\Delta t$ correspond to the spatial and temporal discretization steps, respectively. These values are numerically determined and defined on an equidistant (uniform) grid as $\Delta x = x_i-x_{i-1}$ and $\Delta t = t_j-t_{j-1}$.

The following temperature function is considered:
\begin{equation}
\label{upartida}
T(x,t)=
\begin{cases}
T_1(x,t), & \quad (x,t) \in [0,l_1] \times [0,t_\infty],  \\
T_2(x,t), & \quad (x,t) \in [l_1,l_2] \times [0,t_\infty],  \\
\quad \vdots & \qquad \qquad\qquad\vdots \\
T_M(x,t), & \quad (x,t) \in [l_{M-1},L] \times [0,t_\infty],  \\
\end{cases}
\end{equation}

In order to find the numerical solution of the heat transfer problem studied, equations \eqref{Ec_Parab}-\eqref{Cond_Inicial} are discretized under this scheme. Hence, the following algebraic system can be deduced:
\begin{equation}
\label{Discret}
\begin{cases} 
T^m_{i,j+1}=\zeta_{m,1}\, T^m_{i+1,j}+\zeta_{m,2}\, T^m_{i,j} + \zeta_{m,3}\, T^m_{i-1,j}+ P^m_{i,j}, &  i=n_{l_{m-1}+1},...,n_{l_{m}-1},  j=2,...,J,\\
T^m_{i,j}=T^m_i,  &  i=n_{l_{m-1}},...,n_{l_m}, \,\, j=1, \\
T^1_{i,j}=\epsilon_1 \, T^1_{i+1,j} ,  &  i=1, \, j=2,...,J,\\ 
T^M_{i,j}=\epsilon_M \, T^M_{i-1,j},  &  i=n_{l_M}, \, j=2,...,J, \\
T^m_{i,j}=\upsilon_{m,1} \, T^m_{i-1,j}+ \upsilon_{m,2} \, T^{m+1}_{i+1,j},  &  i=n_{l_m}, \, j=2,...,J,\\
T^{m+1}_{i,j}=\iota_{m,1} \, T^m_{i-1,j}+ \iota_{m,2}\, T^{m+1}_{i+1,j},  &  i=n_{l_m}, \, j=2,...,J,\\
\end{cases}
\end{equation}
where

\begin{equation}
\label{OtroAux}
\begin{cases} 
\zeta_{m,1}=\dfrac{\alpha_m \,\Delta t }{(\Delta x)^2}-\dfrac{\beta_m \,\Delta t }{2 \, \Delta x}, \,\,\,
\zeta_{m,2}=1- 2 \dfrac{\alpha_m \,\Delta t }{(\Delta x)^2}+\nu_m \, \Delta t, \,\,\,
\zeta_{m,3}=\dfrac{\alpha_m \,\Delta t }{(\Delta x)^2}+\dfrac{\beta_m \,\Delta t }{2 \, \Delta x},  \vspace{0.2cm} \\
P^m_{i,j}=s^m_{i,j} \, \Delta t, \,\,\,
\epsilon_1=\dfrac{1}{1+\Delta x \, \Pi_1},  \,\,\,  \epsilon_M=\dfrac{1}{1-\Delta x \, \Pi_M },
\,\,\, \Pi_1=\dfrac{\beta_1}{\alpha_1}+\dfrac{h_1}{\kappa_1}, \vspace{0.2cm} \\ \Pi_M= \dfrac{\beta_M}{\alpha_M}-\dfrac{h_M}{\kappa_M},\,\,\, \Omega_m=\dfrac{R_m}{\Delta x}, \,\,\, Z_m= \dfrac{\beta_m}{\alpha_m} \, \Delta x,    \,\,\,\upsilon_{m,1}=\dfrac{\kappa_m + \kappa_{m+1} \, \Omega_m (1+ Z_{m+1})}{\Lambda_m},  \vspace{0.2cm} \\\upsilon_{m,2}=\dfrac{\kappa_{m+1}}{\Lambda_m}, \,\,\,
\iota_{m,1}= (\Omega_m+1) \, \upsilon_{m,1}-\Omega_m,   \,\,\,\iota_{m,2}=(\Omega_m+1) \, \upsilon_{m,2}, \vspace{0.2cm} \\ \Lambda_m= \kappa_m  \, (1-Z_m)+ \kappa_{m+1}  \,(1+ Z_{m+1}) \, (\Omega_m+1).
\end{cases}
\end{equation}

The convergence and stability conditions of this method are documented in the bibliography \cite{Morton05}, 
where for the problem treated here takes the form

\begin{equation}
\label{cond_est}
\left(\dfrac{\beta_m \,\Delta t }{2 \, \Delta x}\right)^2<2 \, \dfrac{\alpha_m \,\Delta t }{(\Delta x)^2}< 1, \qquad \forall m=1,...,M. 
\end{equation}

Under these conditions, it is guaranted a precision of first order in time and of second order in space for the algebraic problem 
\eqref{Discret}-\eqref{OtroAux}.

\section{Numerical Example}

A non-parallel computational scheme was implemented in MATLAB. The simulated results are obtained within a few minutes on a machine equipped with a 4 GHz Intel(R) Core(TM) i7-6700K processor.

This study addresses a single representative example to demonstrate that the numerical method is both stable and convergent. As evidenced in the literature, similar configurations are expected to yield analogous results. Furthermore, the findings presented in this article are applicable to any type of material, provided that the specified conditions and assumptions are met. This applicability arises from the fact that both the analytical and numerical solutions depend solely on the thermal conductivity and diffusivity coefficients of the materials.

Heat transfer is modeled in a four-layer material composed of nickel, aluminum, copper, and silver (Ni-Al-Cu-Ag), immersed in air at normal pressure. The convective heat coefficients $h_1$ and $h_4$ are determined according to \cite{Umbricht20conv}. The thermal parameters of the materials are taken from Table \ref{Prop_Termicas}.

\begin{xmpl}
\label{example1}

For this example the following parameters are considered:
$M=4$, $L=1 \, m$ , $l_1=0.25 \, m$, $l_2=0.50 \, m$, $l_3=0.75 \, m$, $t_\infty=72000 \, s=20 \, h$, $\beta_1=\beta_2=\beta_3=\beta_4=0.02 \, m/s$ , $\nu_1=\nu_2=\nu_3=\nu_4=-0.0001 \,\, 1/s$, $R_1=R_2=R_3=0.1 \, m$ .

The initial condition is null $T_{1,0}(x)=T_{2,0}(x)=T_{3,0}(x)=T_{4,0}(x)=0$ and the heat generation source $s(x,t)$ is a continuous and differentiable function. For $t \in [0, t_\infty]$ and $m = 1, 2, 3, 4$ it is given by:
\begin{equation}
\label{fuente}
s(x,t)=s_m(x,t)= \dfrac{(M+m) \dfrac{^{\circ}C}{m^2 \, s} \, }{ t_\infty^2} \, \, (x-l_{m-1}) \, (x-l_m)\, t\, (t-t_\infty),  x \in [l_{m-1},l_m],  
\end{equation}
equivalently, for $t \in [0, t_\infty]$
\begin{equation}
\label{fuente2} s(x,t)=
\begin{cases} 
s_1(x,t)= \dfrac{5 \dfrac{^{\circ}C}{m^2 \, s} \, }{ t_\infty^2} \, x \, (x-l_1)\, t\, (t-t_\infty), &  x \in [0,l_1],  \vspace{0.35cm} \\
s_2(x,t)= \dfrac{6 \dfrac{^{\circ}C}{m^2 \, s} \, }{t_\infty^2} \, (x-l_1)\,(x-l_2)\, t\, (t-t_\infty), &  x \in [l_1,l_2],   
\vspace{0.35cm} \\
s_3(x,t)= \dfrac{7 \dfrac{^{\circ}C}{m^2 \, s} \, }{t_\infty^2} \, (x-l_2)\,(x-l_3)\, t\, (t-t_\infty), &  x \in [l_2,l_3],  
\vspace{0.35cm} \\
s_3(x,t)= \dfrac{8 \dfrac{^{\circ}C}{m^2 \, s} \, }{t_\infty^2} \, (x-l_3)\,(x-L)\, t\, (t-t_\infty), &  x \in [l_3,L].
\end{cases}
\end{equation}
\end{xmpl}

\begin{figure}[!h]
\begin{center}
\includegraphics[width=0.495\textwidth]{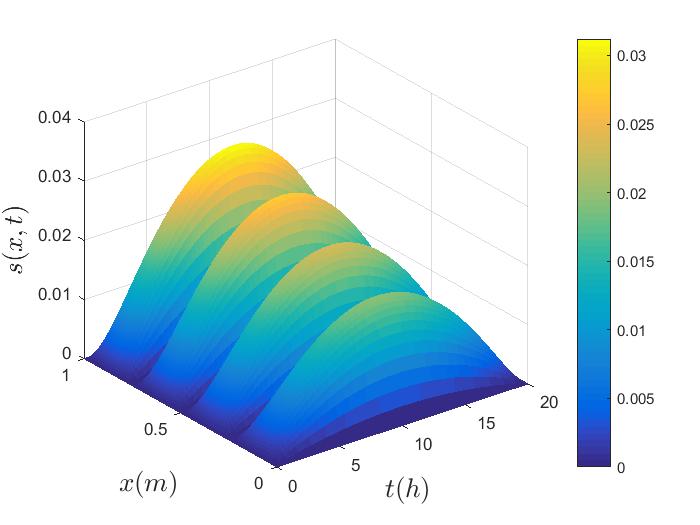}
\includegraphics[width=0.495\textwidth]{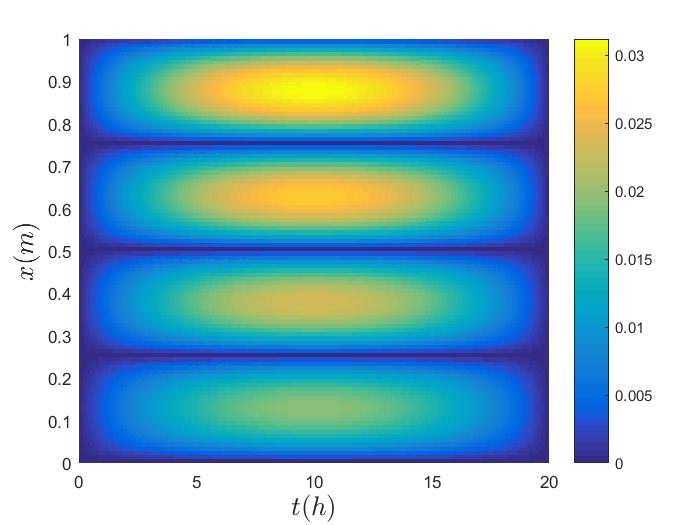}
\vspace{-0.5cm} 
\caption{Heat source.}
\vspace{-0.5cm}
\label{Heat_source}
\end{center}
\end{figure}

This type of source is interesting because it models the heating from the
center of each layer to its edges, where heat generation is zero. Similar to
what happens when heat is delivered to a system through a point thermal
source. In figure \ref{Heat_source} it can be seen that the maximum heating of each layer
goes from 0.020 $^{\circ}C/s$ to 0.030 $^{\circ}C/s$ approximately. These maximum sources
of heat generation occur at $x = 0.125 \, m$, $t = 10 \, h$ for the first layer, at
$x = 0.375 \, m$, $t = 10 \, h$ for the second layer, at $x = 0.625 \, m$, $t = 10 \, h$ in the
third and at $x = 0.875 \,m$, $t = 10 \,h$ for the fourth.

\begin{figure}[!h]
\begin{center}
\includegraphics[width=0.495\textwidth]{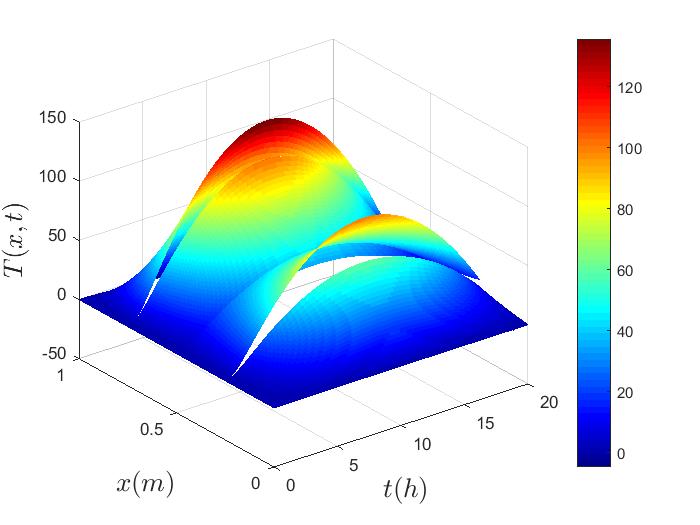}
\includegraphics[width=0.495\textwidth]{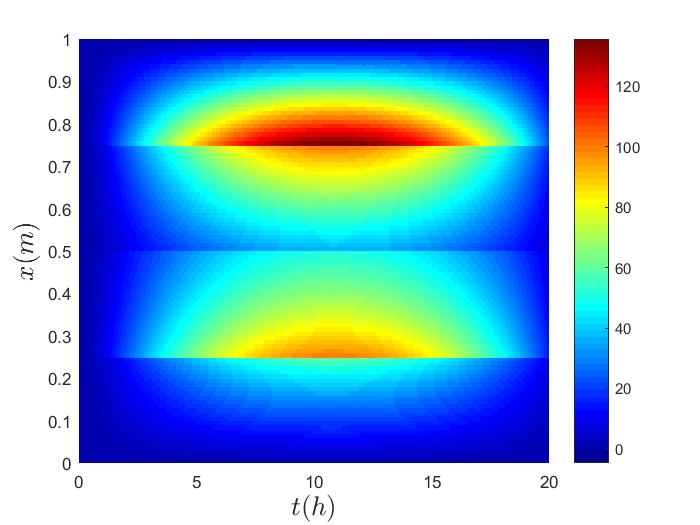}
\vspace{-0.5cm} 
\caption{Temperature distribution for Ni-Al-Cu-Ag}
\vspace{-0.5cm}
\label{Distribucion de temperatura}
\end{center}
\end{figure}

In figure \ref{Distribucion de temperatura} the spatio-temporal temperature profile is observed. The
temperature discontinuities due to the thermal jump at $l_1 = 0.25 \,m$, $l_2 =
0.50 \, m$ and $l_3 = 0.75 \, m$ are displayed. Furthermore, it can be seen that the
maximum temperature of each layer is reached at $t = 10 \, h$ which is directly
related to the nature of the thermal source.

On the other hand, it is clearly seen that the temperature reached by the first layer is lower than that reached by the second layer, the temperature of the second layer is lower than that of the third, and the temperature of the third layer is lower than the temperature reached by the last layer. This is because silver is a more conductive and diffusive material than copper, which in turn is more conductive and diffusive than aluminum, which is more conductive and diffusive than nickel. These results are consistent with the physics of the problem.

\section{Conclusions}

This paper provides a theoretical analysis of a one-dimensional heat transfer problem in a layered body consisting of $m$ layers. The analysis encompasses diffusion, advection, and internal heat generation or loss, which varies linearly with temperature in each layer, as well as heat generation from external sources. Additionally, the thermal resistance at the interfaces between the different materials and general convective boundary conditions are taken into account.

An analytical solution is derived for the problem through the application of dimensionless variable transformations and differential equation techniques, including separation of variables, Fourier methods, and the superposition principle. The analysis reveals that the associated eigenvalue equation possesses infinitely many solutions, and an orthogonality condition is established. The analytical solution is demonstrated to be consistent with findings in previous literature for simpler cases, thereby validating the methodology employed in this study.

Moreover, a convergent finite difference method is introduced, which incorporates a tailored approach at the interfaces, resulting in a mixed finite difference scheme. This method effectively models the problem, providing valuable insights into temperature profiles and the behavior of materials under varying conditions. The numerical results align with the physical expectations of the problem. Specifically, the spatiotemporal temperature profiles exhibit a functional form similar to that of the source, and the response of different materials corresponds with their diffusivity and thermal conductivity: materials with higher diffusivity show a faster increase in temperature, while those with greater thermal conductivity achieve higher temperatures.

\section*{Acknowledgments}

The first author thanks the project \textit{``Problemas de transferencia de calor en materiales multicapa y determinación de la fuente en ecuaciones parabólicas completas''} from Universidad Austral, Rosario, Argentina.

\begin{app}
\section*{Solution to the homogeneous system}
\label{SHS}
The homogeneous system associated with \eqref{Sist_w1}-\eqref{Aux1}, i.e., without source terms, is addressed using the method of separation of variables.   

It is assumed that there exist functions $f_{m,n} \in C^2 (\bar{l}_{m-1},\bar{l}_m)$ and $g_n \in C^1 (0,+\infty)$ such that 
\begin{equation}
\label{SepVar1}
\Theta_m^H(y,\tau)= \ds \sum_{n=1}^{\infty} {f_{m,n} (y) \, g_n(\tau)}, \qquad  (y,\tau) \in (\bar{l}_{m-1},\bar{l}_m) \times \R^+.
\end{equation}

By substituting \eqref{SepVar1} in the homogeneous system associated, it can be shown that $g_n (\tau) = K_n \, \exp(-\lambda_n^2 \, \tau)$, where $\lambda_n$ are the eigenvalues and $K_n$ is a sequence associated with the initial temperature value. In addition, the functions $f_{m,n}$ for $m=1,2,...,M-1$ satisfy 
\begin{equation}
\label{fs}
\begin{cases} 
\bar{\alpha}_m \, f''_{m,n}(y)+\psi_m \,f_{m,n}(y)= - \lambda_n^2 \,f_{m,n}(y), \qquad & y \in (\bar{l}_{m-1},\bar{l}_m), \\ 
f'_{1,n}(y)=  \bar{{Bi}}_1 \, f_{1,n}(y) , \, & y=0 , \\
f'_{M,n}(y)=  \bar{{Bi}}_M \, f_{M,n}(y) , \, & y=1 , \\
f_{m+1,n}(y) = \phi_m \, f_{m,n}(y) + \mu_m \, f'_{m,n}(y) , \, & y=\bar{l}_m, \\
f'_{m+1,n}(y) = \eta_m \, f_{m,n}(y) + \varphi_m \, f'_{m,n}(y) , \, & y=\bar{l}_m, 
\end{cases}
\end{equation}
yielding
\begin{equation}
\label{fs2}
f_{m,n}(y)= A_{m,n}  \cos (\omega_{m,n} \, y) + B_{m,n}  \sin (\omega_{m,n}\, y), \qquad  y \in [\bar{l}_{m-1},\bar{l}_m].
\end{equation}

Then, the solutions of the homogeneous system are written as
\begin{equation}
\label{Sol}
\Theta_m^H(y,\tau)= \ds \sum_{n=1}^{\infty} { K_n \left[A_{m,n}  \cos (\omega_{m,n} \, y) + B_{m,n}  \sin (\omega_{m,n}\, y) \right] \exp(-\lambda_n^2 \, \tau)}, 
\end{equation}
where $\omega_{m,n}$ are the spatial eigenvalues, which are given by
\begin{equation}
\label{Omegas}
\omega_{m,n}=\omega_{m,n}(\lambda_n)=\sqrt{\dfrac{\lambda_n^2+\psi_m}{\bar{\alpha}_m}}=\sqrt{\dfrac{\lambda_n^2+\bar{\nu}_m-\bar{\alpha}_m \chi_m^2}{\bar{\alpha}_m}}=\sqrt{\dfrac{\lambda_n^2+\bar{\nu}_m- \frac{Pe_m^2}{4\bar{\alpha}_m}}{\bar{\alpha}_m}}.
\end{equation}

Now, \(A_{m,n}\), \(B_{m,n}\) with \(m=1,...,M-1\), and \(\lambda_n\) in \eqref{Sol}-\eqref{Omegas} will be determined. For this, the boundary and interface conditions from \eqref{fs} are used. Additionally, it is assumed that the associated homogeneous system has a non-trivial solution. Algebraic operations are performed and the following expressions are obtained $A_{1,n}=1$, $B_{1,n}=\frac{\bar{{Bi}}_1}{\omega_{1,n}}$ and
\begin{equation}
\label{An}
\begin{split}
A_{m+1,n}&= \dfrac{\sin(\omega_{m,n} \, \bar{l}_m)}{\cos(\omega_{m+1,n} \, \bar{l}_m)} \left(\phi_m \, B_{m,n}-\mu_m \, \omega_{m,n} \, A_{m,n}\right)
\\  &   + \dfrac{\cos(\omega_{m,n} \, \bar{l}_m)}{\cos(\omega_{m+1,n} \, \bar{l}_m)} \left(\phi_m \, A_{m,n} +\mu_m \, \omega_{m,n} \, B_{m,n}\right) - \tan (\omega_{m+1,n} \, \bar{l}_m) \, B_{m+1,n}   
\end{split}
\end{equation}
\begin{equation}
\label{Bn}
\begin{split}
 B_{m+1,n} = & \sin(\omega_{m+1,n} \, \bar{l}_m) \left[\sin(\omega_{m,n} \, \bar{l}_m) \left(\phi_m \, B_{m,n} - \mu_m \, \omega_{m,n} \, A_{m,n}\right)  \right] \\
 +& \sin(\omega_{m+1,n} \, \bar{l}_m) \left[ \cos(\omega_{m,n} \, \bar{l}_m) \left(\phi_m \, A_{m,n}+ \mu_m \, \omega_{m,n} \, B_{m,n}\right)\right] \\
 +& \dfrac{\cos(\omega_{m+1,n} \, \bar{l}_m)}{\omega_{m+1,n}}  \left[\sin(\omega_{m,n} \, \bar{l}_m)\left(\eta_m \, B_{m,n} - \varphi_m \, \omega_{m,n} \, A_{m,n}\right) \right] \\
+& \dfrac{\cos(\omega_{m+1,n} \, \bar{l}_m)}{\omega_{m+1,n}}  \left[ \cos(\omega_{m,n} \, \bar{l}_m) \left(\eta_m \, A_{m,n}+\varphi_m \, \omega_{m,n} \, B_{m,n} \right)\right].
\end{split}
\end{equation}

The eigenvalues $\lambda_n$ are the infinite solutions of the transcendental eigenvalue equation given by:
\begin{equation}
\label{Tang}
\tan (\omega_{M,n}(\lambda_n))=\dfrac{\omega_{M,n} (\lambda_n) \, B_{M,n}- \bar{{Bi}}_M \, A_{M,n}}{\bar{{Bi}}_M \, B_{M,n}+\omega_{M,n}(\lambda_n)\,  A_{M,n}},
\end{equation}
with $A_{M,n}$ and $B_{M,n}$ given by \eqref{An} and \eqref{Bn} (specializing in $m+1=M$) respectively.

\end{app}

\begin{app}
\section*{Study of eigenvalues}
\label{Autovalores}

Using the principle of superposition, the solution to the problem of interest can be expressed as an infinite series. This approach assumes that the set of solutions to the transcendental eigenvalue equation is countably infinite, meaning there are infinitely many eigenvalues $\lambda_n$ that satisfy the equation.  

In this work, only real eigenvalues will be considered since we assume that there is no overheating or thermal runaway in the thermal process under study. Imaginary eigenvalues of $\lambda_n$ would cause an exponential increase in temperature over prolonged times, which corresponds to thermal runaway \cite{Jain21}.

This section will discuss the existence of infinitely many real solutions $\lambda_n$ of the eigenvalue equation. This equation is given by:
\begin{equation}
\label{Tangbis}
\tan (\omega_{M,n}(\lambda_n))=\dfrac{\omega_{M,n} (\lambda_n) \, B_{M,n}- \bar{{Bi}}_M \, A_{M,n}}{\bar{{Bi}}_M \, B_{M,n}+\omega_{M,n} (\lambda_n)\,  A_{M,n}},
\end{equation}
where
\begin{equation}
\label{An2}
\begin{split}
A_{M,n}&= \dfrac{\sin(\omega_{M-1,n} \, \bar{l}_{M-1})}{\cos(\omega_{M,n} \, \bar{l}_{M-1})} \left(\phi_{M-1} \, B_{M-1,n}-\mu_{M-1} \, \omega_{M-1,n} \, A_{M-1,n}\right)
\\  &   + \dfrac{\cos(\omega_{M-1,n} \, \bar{l}_{M-1})}{\cos(\omega_{M,n} \, \bar{l}_{M-1})} \left(\phi_{M-1} \, A_{M-1,n} +\mu_{M-1} \, \omega_{M-1,n} \, B_{M-1,n}\right) \\  & - \tan (\omega_{M,n} \, \bar{l}_{M-1}) \, B_{M,n}, 
\end{split}
\end{equation}
\begin{equation}
\label{Bn2}
\begin{split}
 B_{M,n} = & \sin(\omega_{M,n} \, \bar{l}_{M-1}) \left[\sin(\omega_{M-1,n} \, \bar{l}_{M-1}) \left(\phi_{M-1} \, B_{M-1,n} - \mu_{M-1} \, \omega_{M-1,n} \, A_{M-1,n}\right)  \right] \\
 +& \sin(\omega_{M,n} \, \bar{l}_{M-1}) \left[ \cos(\omega_{M-1,n} \, \bar{l}_{M-1}) \left(\phi_{M-1} \, A_{M-1,n}+ \mu_{M-1} \, \omega_{M-1,n} \, B_{M-1,n}\right)\right] \\
 +& \dfrac{\cos(\omega_{M,n} \, \bar{l}_{M-1})}{\omega_{M,n}}  \left[\sin(\omega_{M-1,n} \, \bar{l}_{M-1})\left(\eta_{M-1} \, B_{M-1,n} - \varphi_{M-1} \, \omega_{M-1,n} \, A_{M-1,n}\right) \right] \\
+& \dfrac{\cos(\omega_{M,n} \, \bar{l}_{M-1})}{\omega_{M,n}}  \left[ \cos(\omega_{M-1,n} \, \bar{l}_{M-1}) \left(\eta_{M-1} \, A_{M-1,n}+\varphi_{M-1} \, \omega_{M-1,n} \, B_{M-1,n} \right)\right].
\end{split}
\end{equation}
with  $A_{1,n}=1$, $B_{1,n}=\dfrac{\bar{{Bi}}_1}{\omega_{1,n}}$ and
\begin{equation}
\label{Omegasbis}
\omega_{m,n}=\omega_{m,n}(x)=\sqrt{\dfrac{x^2+\psi_m}{\bar{\alpha}_m}}=\sqrt{\dfrac{x^2+\bar{\nu}_m-\bar{\alpha}_m \chi_m^2}{\bar{\alpha}_m}}=\sqrt{\dfrac{x^2+\bar{\nu}_m- \frac{Pe_m^2}{4\bar{\alpha}_m}}{\bar{\alpha}_m}}
\end{equation}
and
\begin{equation}
\label{Aux12}
\begin{cases} 
\psi_m= \bar{\nu}_m - \bar{\alpha}_m \, \chi_m^2, \,\,\, 
\bar{{Bi}}_1 = {{Bi}_1}^* -\chi_1, \,\,\,   \bar{{Bi}}_M = {{Bi}_M}^*-\chi_M, \,\,\,
\phi_m= \xi_m \,\delta_m, \vspace{0.2cm}\\
\mu_m= \xi_m\,\bar{R}_m , \,\,\, \varphi_m=\xi_m \, \left(\sigma_m - \bar{R}_m \,\chi_{m+1} \right), \,\,\, \eta_m=\xi_m \, \left(\gamma_m+ \sigma_m \,\chi_m - \delta_m\,\chi_{m+1}   \right), \vspace{0.2cm}\\
\xi_m= \exp\left(\, \bar{l}_m \, (\chi_m-\chi_{m+1})\right), \,\,\, \delta_m=1+ \bar{R}_m \, \chi_m, \,\,\, \chi_m= \dfrac{Pe_m}{2 \,\bar{\alpha}_m}, \,\,\,	{{Bi}_1}^*=\dfrac{Pe_1}{\bar{\alpha}_1}+\dfrac{{Bi}_1}{\bar{\kappa}_1},  \vspace{0.2cm} \\
{{Bi}_M}^*= Pe_M - {Bi}_M, \,\,\, \gamma_m= \dfrac{Pe_{m+1}}{\bar{\alpha}_{m+1}}- \dfrac{Pe_{m}}{\bar{\alpha}_{m}}\, \dfrac{\bar{\kappa}_m}{\bar{\kappa}_{m+1}}, \,\,\, 
\sigma_m=\dfrac{\bar{\kappa}_m}{\bar{\kappa}_{m+1}}+ \bar{R}_m \, \dfrac{Pe_{m+1}}{\bar{\alpha}_{m+1}}, \vspace{0.2cm}\\
 \bar{l}_m=\dfrac{l_m}{L}, \,\,\, \bar{R}_m=\dfrac{R_m}{L}, \,\,\,  \bar{\alpha}_m=\dfrac{\alpha_m}{\alpha_M}, \,\,\, {Pe}_m=\dfrac{L}{\alpha_M} \, \beta_m,  \,\,\,
 \bar{\nu}_m=\dfrac{L^2}{\alpha_M} \,\nu_m, \,\,\, \bar{\kappa}_m=\dfrac{\kappa_m}{\kappa_M}, \vspace{0.2cm} \\ 
{Bi}_1=\dfrac{L}{\kappa_M} \, h_1, \,\,\, {Bi}_M=\dfrac{L}{\kappa_M} \, h_M.
\end{cases}
\end{equation}

Since the equation \eqref{Tangbis}-\eqref{Aux12} is transcendental, it is not possible to obtain solutions. On the other hand, analytically proving that this equation has infinite solutions for the general case is a difficult task due to the complexity of the equation. However, this fact can be verified numerically for each particular case of interest.

If we denote by:
\begin{equation}
\label{OtroAuxbis}
r(x)=\dfrac{\omega_{M,n} (x) \, B_{M,n}(x)- \bar{{Bi}}_M \, A_{M,n}(x)}{\bar{{Bi}}_M \, B_{M,n}(x)+\omega_{M,n} (x)\,  A_{M,n}(x)}, \qquad 
q(x)=\tan (\omega_{M,n}(x)), 
\end{equation}
showing that the eigenvalue equation has infinitely many real solutions boils down to seeing that the functions $r(x)$ and $q(x)$ have infinitely many intersections. As an example, we will observe this in two particular cases for four-layer bodies.
\setcounter{case}{0}
\setcounter{table}{0}
\begin{case}
\label{case1}
The heat transfer problem in a $Al-Cu-Fe-Ni$ four-layer body, is considered. 
\end{case}

\begin{case}
\label{case2}
The heat transfer problem in a $Pb-Al-Ni-Ag$ four-layer body, is considered.
\end{case}

The thermal parameters of the materials are taken from \cite{Cengel07} and summarized in table \ref{Prop_Termicas}. The physical parameters used for the examples in case \ref{case1} and case \ref{case2} are listed in table \ref{PPC1} and table \ref{PPC2}, respectively.

\setcounter{table}{0}
\begin{table}[h!]
\begin{center}
{\begin{tabular}{lccc} \toprule
Materials & Symbol & $\alpha^2 \left(\times 10^{4}\right) \, \left[m^{2}/s\right] $  & $\kappa \, \left[W/m^{\circ}C\right] $\\ \midrule
Lead  &     Pb       &                 0.23673                                  &              35                      \\        
Iron   & Fe         &                 0.20451                                  &              73                      \\ 
Nickel  & Ni         &                  0.22663                                  &              90                      \\ 
Aluminium   & Al         &                 0.84010                                  &              204                     \\ 
Copper    &  Cu         &                 1.12530                                  &              386                     \\ 
Silver   &  Ag         &                 1.70140                                  &              419                     \\ \bottomrule
\end{tabular}}  
\end{center}
\vspace{-0.5cm}
\caption{Thermal properties of different materials.}
\label{Prop_Termicas}
\end{table}

\begin{table}[h!]
\begin{center}
\begin{minipage}{0.45\textwidth}
\centering
\begin{tabular}{lc} \toprule
Parameters                                                       &       Values   \\ \midrule
$L \, \left[m\right] $                                           &       1       \\ 
$l_1 \, \left[m\right] $                                           &       0.2      \\ 
$l_2 \, \left[m\right] $                                           &       0.5      \\ 
$l_3 \, \left[m\right] $                                           &       0.8      \\ 
$h_1 \, \left[W/m^2\,^{\circ}C\right] $                            &       12        \\  
$h_4 \, \left[W/m^2\,^{\circ}C\right] $                            &       10        \\  
$\beta_1 \, \left[m/s\right] $                                     &       0.001       \\   
$\beta_2 \, \left[m/s\right] $                                     &       0.002       \\ 
$\beta_3 \, \left[m/s\right] $                                     &       0.003       \\   
$\beta_4 \, \left[m/s\right] $                                     &       0.004       \\ 
$\nu_1 \, \left[1/s\right] $                                     &       10       \\   
$\nu_2 \, \left[1/s\right] $                                     &       15       \\ 
$\nu_3 \, \left[1/s\right] $                                     &       10       \\   
$\nu_4 \, \left[1/s\right] $                                     &       15       \\ 
$R_1\, \left[m\right] $                                                  &       0.04        \\
$R_2\, \left[m\right] $                                                  &       0.05       \\$R_3\, \left[m\right] $                                                  &       0.06        \\   \bottomrule
\end{tabular}
\caption{Physical parameters of case \ref{case1}.}
\vspace{-0.5cm}
\label{PPC1}
\end{minipage}
\hspace{0.05\textwidth}
\begin{minipage}{0.45\textwidth}
\centering
\begin{tabular}{lc} \toprule
Parameters                                                       &       Values   \\ \midrule
$L \, \left[m\right] $                                           &       2       \\ 
$l_1 \, \left[m\right] $                                           &       0.7      \\ 
$l_2 \, \left[m\right] $                                           &       1.2      \\ 
$l_3 \, \left[m\right] $                                           &       1.6      \\ 
$h_1 \, \left[W/m^2\,^{\circ}C\right] $                            &       10        \\  
$h_4 \, \left[W/m^2\,^{\circ}C\right] $                            &       12        \\  
$\beta_1 \, \left[m/s\right] $                                     &       0.003       \\   
$\beta_2 \, \left[m/s\right] $                                     &       0.001       \\ 
$\beta_3 \, \left[m/s\right] $                                     &       0.002       \\   
$\beta_4 \, \left[m/s\right] $                                     &       0.004       \\ 
$\nu_1 \, \left[1/s\right] $                                     &       15       \\   
$\nu_2 \, \left[1/s\right] $                                     &       20       \\ 
$\nu_3 \, \left[1/s\right] $                                     &       15       \\   
$\nu_4 \, \left[1/s\right] $                                     &       20       \\ 
$R_1\, \left[m\right] $                                                  &       0.07        \\
$R_2\, \left[m\right] $                                                  &       0.06       \\$R_3\, \left[m\right] $                                                  &       0.05        \\   \bottomrule
\end{tabular}
\caption{Physical parameters of case \ref{case2}.}
\vspace{-0.5cm}
\label{PPC2}
\end{minipage}
\end{center}
\end{table}

From Fig.\ref{inf_sol} you can see the intercessions for $\lambda_n \in (-200,200)$. It can be inferred, for both cases, that the functions $q(x)$ and $r(x)$ will have, effectively, infinite intersections.

\begin{figure}[!h]
\begin{center}
\includegraphics[width=0.495\textwidth]{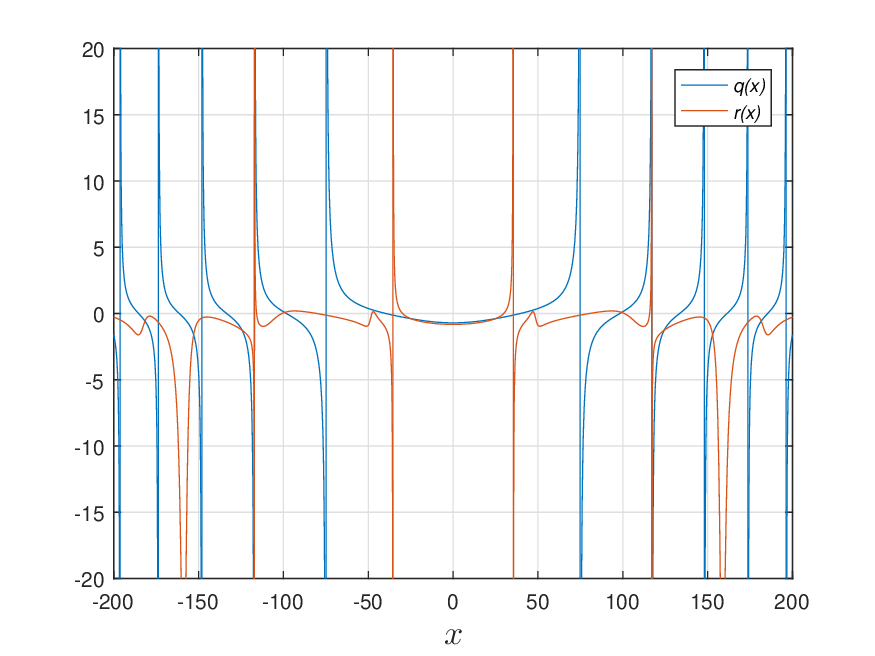}
\includegraphics[width=0.495\textwidth]{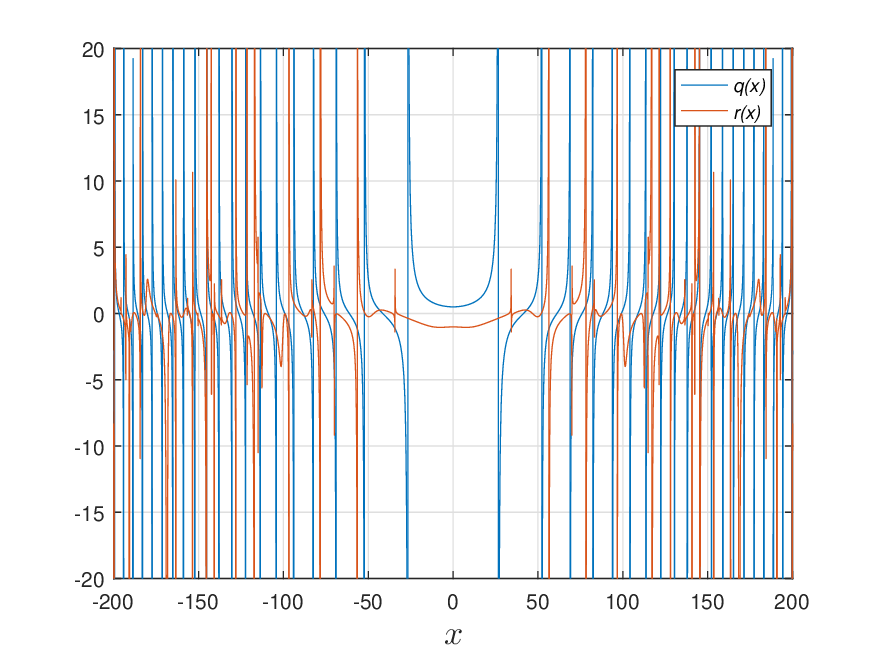}
\vspace{-0.5cm} 
\caption{Scheme of intersection of the functions $q(x)$ and $r(x)$. To the left for case \ref{case1} and to the right for case \ref{case2}.}
\vspace{-0.5cm}
\label{inf_sol}
\end{center}
\end{figure}

\end{app}

\begin{app}

\section*{Study of the orthogonality relationship} \label{Ortogonalidad}

In this section, we will derive the orthogonality condition, or principle, for this problem. This result is necessary to determine the sequence $K_n$ in \eqref{Soluciongeneralhomogeneo}. As shown in \eqref{fs}, for two numbers $n$ and $j$, the functions $f_{m,n}$ and $f_{m,j}$ for all $m=1,2,...,M$ must satisfy:
\begin{equation}
\label{f1nj}
\begin{cases} 
\bar{\alpha}_m \, f''_{m,n}(y)+\psi_m \,f_{m,n}(y)= - \lambda_n^2 \,f_{m,n}(y), \quad & y \in (\bar{l}_{m-1},\bar{l}_m), \\ 
\bar{\alpha}_m \, f''_{m,j}(y)+\psi_m \,f_{m,j}(y)= - \lambda_j^2 \,f_{m,j}(y), \quad & y \in (\bar{l}_{m-1},\bar{l}_m)
\end{cases}
\end{equation}

Multiply the first equation of \eqref{f1nj} by $f_{m,j}$ and the second by $f_{m,n}$. This gives rise to,
\begin{equation}
\label{f1njbis}
\begin{cases} 
\bar{\alpha}_m \, f''_{m,n}(y) \,f_{m,j}(y)+\psi_m \,f_{m,n}(y)\,f_{m,j}(y)= - \lambda_n^2 \,f_{m,n}(y)\,f_{m,j}(y), \, & y \in (\bar{l}_{m-1},\bar{l}_m),   \\ 
\bar{\alpha}_m \, f''_{m,j}(y)\,f_{m,n}(y)+\psi_m \,f_{m,j}(y)\,f_{m,n}(y)= - \lambda_j^2 \,f_{m,j}(y)\,f_{m,n}(y), \, & y \in (\bar{l}_{m-1},\bar{l}_m). 
\end{cases}
\end{equation}

The difference of the two expressions of \eqref{f1njbis} is taken, 
\begin{equation}
\label{fnjbis}
\bar{\alpha}_m \, \left[f''_{m,n}(y) \,f_{m,j}(y)-f''_{m,j}(y)\,f_{m,n}(y)\right]= (\lambda_j^2- \lambda_n^2) \,f_{m,n}(y)\,f_{m,j}(y), \quad y \in (\bar{l}_{m-1},\bar{l}_m).
\end{equation}

Letting 
 $\Psi_1=1$ and
\begin{equation}
\label{Psim}
\Psi_{m}=\dfrac{1}{\ds \prod_{i=1}^{m-1} N_i}= \dfrac{1}{\ds \prod_{i=1}^{m-1}  (\varphi_i \, \phi_i-\eta_i \, \mu_i)}, \qquad m=2,...,M,
\end{equation}
the expression \eqref{fnjbis} is conveniently rewritten as follows
\begin{equation}
\label{fnjbis2}
 \Psi_{m} \left[f'_{m,n}(y) \,f_{m,j}(y)-f'_{m,j}(y)\,f_{m,n}(y)\right]' = \dfrac{\Psi_{m}}{\bar{\alpha}_m} \,(\lambda_j^2- \lambda_n^2) f_{m,n}(y)f_{m,j}(y).  
\end{equation}

The equalities \eqref{fnjbis2} are integrated over their respective intervals of definition and then added. This yields,
\begin{equation}
\label{fnjbis3}
(\lambda_j^2- \lambda_n^2) \ds \sum_{m=1}^M \dfrac{\Psi_{m}}{\bar{\alpha}_m} \ds \int_{\bar{l}_{m-1}}^{\bar{l}_{m}}  f_{m,n}(y)f_{m,j}(y) \, dy  
=  \ds \sum_{m=1}^M \Psi_{m} \, \Gamma_{m,j,n} (y) \Big|_{\bar{l}_{m-1}}^{\bar{l}_{m}},
\end{equation}
where
\begin{equation}
\label{DefGamma}
\Gamma_{m,j,n} (y) =f'_{m,n}(y) \,f_{m,j}(y)-f'_{m,j}(y)\,f_{m,n}(y).
\end{equation}

By operating algebraically and using the properties of the function $\Gamma$ (properties \ref{propiedadesGamma}) given in expressions \eqref{Prop4}, we arrive at:
\begin{equation}
\label{fnjbis4}
\begin{split}
&(\lambda_j^2- \lambda_n^2) \ds \sum_{m=1}^M \dfrac{\Psi_{m}}{\bar{\alpha}_m} \ds \int_{\bar{l}_{m-1}}^{\bar{l}_{m}}   f_{m,n}(y)f_{m,j}(y) \, dy  
=  \ds \sum_{m=1}^M \Psi_{m} \, \left(\Gamma_{m,j,n} (\bar{l}_{m})-\Gamma_{m,j,n} (\bar{l}_{m-1})\right)\\
 & =\ds \sum_{m=1}^M \Psi_{m} \, \Gamma_{m,j,n} (\bar{l}_{m})-\ds \sum_{m=1}^M \Psi_{m} \, \Gamma_{m,j,n} (\bar{l}_{m-1})= \Psi_{M} \, \Gamma_{M,j,n} (1)-\Psi_{1} \, \Gamma_{1,j,n} (0)\\
 & + \ds \sum_{m=1}^{M-1} \Psi_{m} \, \Gamma_{m,j,n} (\bar{l}_{m})-\ds \sum_{m=2}^M \Psi_{m} \, N_{m-1} \,\Gamma_{m-1,j,n} (\bar{l}_{m-1})\\
&=\ds \sum_{m=1}^{M-1}[\Psi_{m}-N_m \, \Psi_{m+1}] \,\Gamma_{m,j,n} (\bar{l}_{m})=0.
\end{split}
\end{equation}
What is obtained in \eqref{fnjbis4} allows us to deduce the orthogonality relation given by
\begin{equation}
\label{CondOrtogonalidad}
\ds \sum_{m=1}^M \dfrac{\Psi_{m}}{\bar{\alpha}_m} \ds \int_{\bar{l}_{m-1}}^{\bar{l}_{m}}  f_{m,n}(y)f_{m,j}(y) \, dy  =0, \quad \forall n\neq j.
\end{equation}

\begin{prop} (Properties of the $\Gamma$)
\label{propiedadesGamma}

The function $\Gamma$ defined in \eqref{DefGamma} possesses properties of particular interest in the context of this problem. These properties are useful for deriving the orthogonality condition \eqref{CondOrtogonalidad}, and they are obtained from the boundary and interface conditions of \eqref{fs}.  
\begin{equation}
\label{Prop1}
\begin{split}
\Gamma_{1,j,n} (0)&=f'_{1,n}(0) \,f_{1,j}(0)-f'_{1,j}(0)\,f_{1,n}(0) \\ &=\bar{{Bi}}_1 \, f_{1,n}(0) \,f_{1,j}(0)-
\bar{{Bi}}_1 \, f_{1,j}(0) \,f_{1,n}(0)=0.
\end{split}
\end{equation}
\begin{equation}
\label{Prop2}
\begin{split}
\Gamma_{M,j,n} (1)&=f'_{M,n}(1) \,f_{M,j}(1)-f'_{M,j}(1)\,f_{M,n}(1) \\ &=\bar{{Bi}}_M \, f_{M,n}(1) \,f_{M,j}(1)-
\bar{{Bi}}_M \, f_{M,j}(0) \,f_{M,n}(0)=0.
\end{split}
\end{equation}
\begin{equation}
\label{Prop3}
\begin{split}
&\Gamma_{m,j,n} (\bar{l}_{m-1})=f'_{m,n}(\bar{l}_{m-1}) \,f_{m,j}(\bar{l}_{m-1})-f'_{m,j}(\bar{l}_{m-1})\,f_{m,n}(\bar{l}_{m-1}) \\ =& [\eta_{m-1} \, f_{m-1,n} +\varphi_{m-1} \, f'_{m-1,n}]\Big|_{\bar{l}_{m-1}}
[\phi_{m-1} \, f_{m-1,j} +\mu_{m-1} \, f'_{m-1,j}]\Big|_{\bar{l}_{m-1}}
\\-& [\eta_{m-1} \, f_{m-1,j} +\varphi_{m-1} \, f'_{m-1,j}]\Big|_{\bar{l}_{m-1}}
[\phi_{m-1} \, f_{m-1,n} +\mu_{m-1} \, f'_{m-1,n}]\Big|_{\bar{l}_{m-1}}\\
=&\eta_{m-1} \,\mu_{m-1} [f_{m-1,n} \, f'_{m-1,j}-f_{m-1,j} \, f'_{m-1,n}]\Big|_{\bar{l}_{m-1}} \\
+&\varphi_{m-1} \,\phi_{m-1} [f_{m-1,j} \, f'_{m-1,n}-f_{m-1,n} \, f'_{m-1,j}]\Big|_{\bar{l}_{m-1}} \\
=& (\varphi_{m-1} \,\phi_{m-1}- \eta_{m-1} \,\mu_{m-1} ) [f_{m-1,j} \, f'_{m-1,n}-f_{m-1,n} \, f'_{m-1,j}]\Big|_{\bar{l}_{m-1}} \\
=& N_{m-1} \, \Gamma_{m-1,j,n} (\bar{l}_{m-1}).
\end{split}
\end{equation}

In summary, the properties of the function $\Gamma$ are:

\begin{equation}
\label{Prop4}
\Gamma_{1,j,n} (0)=0, \qquad \Gamma_{M,j,n} (1)=0 , \qquad \Gamma_{m,j,n} (\bar{l}_{m-1})= N_{m-1} \, \Gamma_{m-1,j,n} (\bar{l}_{m-1}).
\end{equation}
\end{prop}
\end{app}

\begin{app}
\section*{Solution to the non-homogeneous system}
\label{SNHS}

To find the solution of the non-homogeneous system of interest \eqref{Sist_w1}-\eqref{Aux1} the Fourier method is used. That is, it is assumed that there are countably infinite sets of time functions $\bar{A}_{m,n}(\tau)$ such that
\begin{equation}
\label{SolNoHomegeneo2}
\Theta_m(y,\tau)= \ds \sum_{n=1}^{\infty} {\bar{A}_{m,n}(\tau) \, f_{m,n} (y) }, \quad  (y,\tau) \in [\bar{l}_{m-1},\bar{l}_m] \times \R^+.
\end{equation}
where $f_{m,n}$ with $m=1,2,...,M$ are defined in \eqref{fs2}. For simplicity, the source functions $\widehat{s}_m(y,\tau)$ in \eqref{Sist_w1} are developed in a series of eigenfunctions.
\begin{equation}
\label{Fuenteenserie}
\widehat{s}_m(y,\tau)= \ds \sum_{n=1}^{\infty}  { S_{m,n}(\tau) \, f_{m,n} (y) }, \quad  (y,\tau) \in [\bar{l}_{m-1},\bar{l}_m] \times \R^+,
\end{equation}
where $S_{m,n}(\tau)$ are defined as follows
\begin{equation}
\label{Gs}
S_{m,n}(\tau)=\dfrac{\ds \int_{\bar{l}_{m-1}}^{\bar{l}_m} \widehat{s}_m(y,\tau) \,  f_{m,n} (y) \, dy}{\ds \int_{\bar{l}_{m-1}}^{\bar{l}_m} \left[f_{m,n} (y)\right]^2 \, dy}.
\end{equation}

Replacing the expressions \eqref{SolNoHomegeneo2}-\eqref{Gs} in the equation \eqref{Sist_w1}, the following countable set of homogeneous ordinary equations is obtained,
\begin{equation}
\label{EDO}
\ds \sum_{n=1}^{\infty} \left[\bar{A}_{m,n}'(\tau) +(\bar{\alpha}_m \, \omega^2_{m,n} - \psi_m) \, \bar{A}_{m,n}(\tau) -S_{m,n}(\tau)\right] =0,
\end{equation}
since the expansions in eigenfunctions of linear system problems have properties similar to those of Fourier series, for the series given in \eqref{EDO} to sum to zero, it is necessary that all their terms be zero. This can be solved by direct integration, which gives rise to:
\begin{equation}
\label{solEDO}
\bar{A}_{m,n}(\tau)=\exp\left((\psi_m-\bar{\alpha}_m \, \omega^2_{m,n})\, \tau\right) \left[K_n + \ds \int_0^\tau S_{m,n} (s) \, \exp\left((\bar{\alpha}_m \, \omega^2_{m,n}-\psi_m)\, s\right)  \,ds\right]. 
\end{equation}

Only $K_n$ remains to be determined. This sequence can be found by imposing the initial conditions of \eqref{Sist_w1} and using the orthogonality condition, which will be detailed in the Appendix \ref{Ortogonalidad}. In this way, it is obtained:
\begin{equation}
\label{CI}
K_n= \dfrac{\ds \sum_{m=1}^{M} \dfrac{\Psi_m}{\bar{\alpha}_m} \ds \int_{\bar{l}_{m-1}}^{\bar{l}_m} \Theta_{m,0}(y)\, f_{m,n}(y) \, dy }{\ds \sum_{m=1}^{M}  \dfrac{\Psi_m}{\bar{\alpha}_m}  \ds \int_{\bar{l}_{m-1}}^{\bar{l}_m}  [f_{m,n}(y)]^2 \, dy }.
\end{equation}

\end{app}

\end{document}